\documentclass[twocolumn,times]{aastex62}

\shorttitle{Uncertainties in Atomic Data}
\shortauthors{Yu, Kashyap, Del Zanna, van Dyk, Stenning, et al.}

\usepackage[T1]{fontenc}
\usepackage{parskip}
\usepackage{latexsym,amsmath,amssymb}
\usepackage{graphicx}
\usepackage{times}
\usepackage{zi4}

\usepackage{color}
\usepackage{soul}
\usepackage{float}

\usepackage{multirow}
\usepackage{enumerate}
\usepackage[figuresright]{rotating}
\usepackage[normalem]{ulem}
\usepackage{array}

\usepackage{bm}
\usepackage{esvect}
\usepackage{bbm}

\usepackage{hyperref}
\hypersetup{
    colorlinks=true,
    linkcolor=blue,
    filecolor=magenta,
    urlcolor=cyan,
}

\definecolor{amber}{rgb}{1.0, 0.49, 0.0}
\definecolor{coralpink}{rgb}{0.97, 0.51, 0.47}
\definecolor{green}{rgb}{0.0, 0.5, 0.0}

\newcommand{\Data}{\mathcal{D}} 

\newcommand{\dataso}{Y} 
\newcommand{\databg}{Z} 
\newcommand{\emiss}{\mathcal{M}}
\newcommand{\datafe}{\mathcal{D}^{\rm Fe}}
\newcommand{\emissfe}{\mathcal{M}^{\rm Fe}}
\newcommand{\datafeso}{Y^{\rm Fe}}
\newcommand{\datafebg}{Z^{\rm Fe}}

\newcommand{\dataox}{\mathcal{D}^{\rm O}}
\newcommand{\emissox}{\mathcal{M}^{\rm O}}
\newcommand{\dataoxso}{Y^{\rm O}}
\newcommand{\dataoxbg}{Z^{\rm O}}

\newcommand{\param}{\theta_{go}} %
\newcommand{\paramso}{\theta_{{\rm S}go}} %
\newcommand{\parambg}{\theta_{{\rm B}go}} %
\newcommand{\paramfe}{\theta^{\rm Fe}}
\newcommand{\paramfeso}{\theta^{\rm Fe}_{S}}
\newcommand{\paramfebg}{\theta^{\rm Fe}_{B}}
\newcommand{\paramox}{\theta^{\rm O}}
\newcommand{\paramoxso}{\theta^{\rm O}_{S}}
\newcommand{\paramoxbg}{\theta^{\rm O}_{B}}
\newcommand{\paramoxfe}{\theta^{O,Fe}}

\newcommand{\Ne}{\mathrm{n}} 
\newcommand{\Te}{\mathrm{T}}

\newcommand{\m}{m}
\newcommand{\rvm}{m}

\newcommand{\ec}{\epsilon}

\newcommand{\eigvalsqrt}{\beta}
\newcommand{\eigvec}{v}
\newcommand{\rv}{r}

\newcommand{\rvfe}{r^{\rm Fe}}
\newcommand{\rvox}{r^{\rm O}}

\makeatletter
\newcommand{\distas}[1]{\mathbin{\overset{#1}{\kern\z@\sim}}}%
\newsavebox{\mybox}\newsavebox{\mysim}
\newcommand{\distras}[1]{%
  \savebox{\mybox}{\hbox{\kern3pt$\scriptstyle#1$\kern3pt}}%
  \savebox{\mysim}{\hbox{$\sim$}}%
  \mathbin{\overset{#1}{\kern\z@\resizebox{\wd\mybox}{\ht\mysim}{$\sim$}}}%
}
\makeatother

\newcommand{\chandra}{{\sl Chandra}}
\newcommand{\ionn}{X}
\newcommand{\OHe}{\ion{O}{7}}
\newcommand{\Fe}{\ion{Fe}{17}}  
\newcommand{\Fexiii}{\ion{Fe}{13}}

\newcommand{\wvl}{{\mathrm w}} 
\newcommand{\ww}{\omega} 
\newcommand{\il}{l} 
\newcommand{\ih}{h} 

\newcommand{\iL}{L} 
\newcommand{\iH}{H}

\newcommand{\arfh}{A_{Igo}(\ww_\ih)}

\newcommand{\arfl}{A_{Igo}(\ww_\il)}

\newcommand{\lrf}{{R_{go}(\wvl | \ww)}}
\newcommand{\lrfl}{{R_{go}(\wvl | \ww_\il)}}
\newcommand{\lrflion}{{R_{go}(\wvl^{X} | \ww_\il)}}

\newcommand{\fx}{{f}}

\newcommand{\volume}{{V}}
\newcommand{\logtemp}{{\log\Te}}

\newcommand{\exptim}{{\mathcal{T}}} 
\newcommand{\estso}{s} 
\newcommand{\estfeso}{s^{\rm Fe}}
\newcommand{\estoxso}{s^{\rm O}}

\newcommand{\estcont}{\kappa} 
\newcommand{\estfecont}{\kappa^{\rm Fe}}
\newcommand{\estoxcont}{\kappa^{\rm O}}
\newcommand{\ctnm}{c}
\newcommand{\ratio}{\eta} 
\newcommand{\acismegpos}{{\tt amp}}
\newcommand{\acismegneg}{{\tt amn}}
\newcommand{\acishegpos}{{\tt ahp}}
\newcommand{\acishegneg}{{\tt ahn}}
\newcommand{\acislegpos}{{\tt alp}}
\newcommand{\acislegneg}{{\tt aln}}

\begin{document}


\title{Effect of Systematic Uncertainties on Density and Temperature Estimates in Coronae of Capella}

\author[0000-0001-9793-6179]{Xixi Yu}
\affil{Statistics Section, Department of Mathematics, Imperial College London, London SW7 2AZ, UK}

\author[0000-0002-3869-7996]{Vinay L.\  Kashyap}
\affil{Center for Astrophysics $|$ Harvard \& Smithsonian, 60 Garden Street, Cambridge, MA 02138, USA}

\author[0000-0002-4125-0204]{Giulio Del Zanna}
\affil{DAMTP, Centre for Mathematical Sciences, University of Cambridge, Wilberforce Road, Cambridge CB3 0WA, UK}

\author[0000-0002-0816-331X]{David~A.~van~Dyk}
\affil{Statistics Section, Department of Mathematics, Imperial College London, London SW7 2AZ, UK}

\author[0000-0002-9761-4353]{David C.\ Stenning}
\affil{Statistics and Actuarial Science, Simon Fraser University, 8888 University Dr, Burnaby, BC V5A 1S6, Canada}

\author{Connor P.\ Ballance}
\affil{Queen's University of Belfast, Belfast, UK}

\author{Harry P.\ Warren}
\affil{Space Science Division, Naval Research Laboratory, Washington, DC 20375, USA}


\begin{abstract}
We estimate the coronal density of Capella using the \OHe\ and \Fe\ line systems in the soft X-ray regime that have been observed over the course of the \chandra\ mission.
Our analysis combines measures of error due to uncertainty in the underlying atomic data with statistical errors in the \chandra\ data
to derive meaningful overall uncertainties on the plasma density of the coronae of Capella.
We consider two Bayesian frameworks. First, the so-called pragmatic-Bayesian approach considers the atomic data and their uncertainties as fully specified and uncorrectable. The fully-Bayesian approach, on the other hand, 
allows the observed spectral data to update the atomic data and their uncertainties, thereby reducing the overall errors on the inferred parameters. 
To incorporate atomic data uncertainties, we obtain a set of atomic data replicates, the distribution of which captures their uncertainty. A principal component analysis of these replicates allows us to represent the atomic uncertainty with a lower-dimensional multivariate Gaussian distribution.  
A $t$-distribution approximation of the uncertainties of a subset of plasma parameters including a priori temperature information, obtained from the temperature-sensitive-only \Fe \ spectral line analysis, is carried forward into the density- and temperature-sensitive \OHe\ spectral line analysis.
Markov Chain Monte Carlo based model fitting is implemented including Multi-step Monte Carlo Gibbs Sampler and Hamiltonian Monte Carlo. 
Our analysis recovers an isothermally approximated coronal plasma temperature of $\approx$5~MK and a coronal plasma density of $\approx$10$^{10}$~cm$^{-3}$, with uncertainties of 0.1 and 0.2~dex respectively.
\end{abstract}



\newpage
\pagenumbering{arabic}


\section{Introduction} \label{sec:intro}

The active binary Capella ($\alpha$ Aur Aa+Ab, G8\,III+G1\,III; distance 13~pc) is the brightest stellar coronal X-ray source besides the Sun.  While slow variations over time scales of months and years are observed, the system is remarkably steady at short time scales; for instance, no flares have been observed to occur on it.  The Capella spectrum is rich in emission lines formed over a broad temperature range, and its modelling is complex as multiple temperature and density components are present.  Nevertheless, its emission measure distribution is dominated by plasma at temperatures in the $\sim$5~MK regime \citep{brickhouse2000coronal,2003A&A...404.1033A,2006ApJ...649..979G}, allowing the coronal modeling to be usefully simplified.  High-quality spectroscopic observations of solar and stellar coronae in the X-ray and extreme ultraviolet, usually combined with high quality atomic data, have a tremendous capability for estimating fundamental plasma parameters like temperature distributions and densities.  In these analyses, it is crucial to account for the reliability of atomic data for interpreting and modeling X-ray observations \citep{kallman2007atomic}, as uncertainties in the atomic data are likely to be the prominent sources of error in the spectral analysis and have widespread implications for spectral models of stars \citep{brickhouse1995new, brickhouse2000coronal}.  Thus, ignoring atomic uncertainty not only can lead to underestimation of the variance of plasma parameters but also can cause serious bias in their estimates.

In this work, we aim to understand the effect of the uncertainties in the atomic data and how these uncertainties in the underlying atomic physics propagate to the determination of plasma parameters. We use Capella as the exemplar object to apply our methods. While a sensible physical interpretation of the Capella coronae is still lacking, empirically its emission is observed to be stable, with slow flux variations over time scales of months, and the emission dominated by a near-isothermal plasma. Capella is well-suited for testing the application of incorporating atomic data errors, both because its emission model can be usefully simplified, and because a great deal of data have been collected on it, thus minimizing the effect of statistical fluctuations. Capella has been widely used as a calibration target with \chandra, accumulating several hundred megaseconds of high-quality high-resolution exposure time with the \chandra\ gratings. With its small temporal variability, long exposures can be combined to obtain a high signal-to-noise ratio (S/N) dataset, which minimizes statistical uncertainty and provides an opportunity to explore systematic effects in the modeling and interpretation of the spectrum. In fact, this portends an issue that will become increasingly prevalent in the coming era of X-ray telescopes with large collecting areas and high spectral resolution. In summary, we focus here on characterizing the effects of uncertainties in atomic emissivities in establishing the temperature and plasma density of the coronae of Capella. We demonstrate a statistically principled method that incorporates atomic data uncertainties to estimate these quantities.

Over the years, there has been wide recognition of the importance of accounting for systematic uncertainties, including instrument calibration \citep[e.g.,][]{drake2006monte,lee2011accounting,xu2014fully,2021AJ....162..254M} and atomic data \citep[e.g.,][]{2010SSRv..157..135F,yu2018incorporating,2021ApJ...908....3H}.  Systematic uncertainties are often ignored due to a lack of robust and principled methods that can handle them, as well as a lack of characterization of their magnitudes.  But ignoring them can introduce biases in the estimation of model parameters and/or result in the underestimation of their variance, and thus lead to erroneous interpretations of astrophysical data.

Incorporating systematic uncertainties is a complex problem.  \citet{drake2006monte} first proposed a bootstrap-type method to account for systematics arising from instrument effective area uncertainties, by relying on a large sample of possible calibration values, and demonstrated that such systematic uncertainty can be propagated to parameter estimates. \citet{lee2011accounting} used a principal component analysis (PCA) to efficiently model such a high-dimensional large calibration sample with a low-dimensional distribution and developed a \textit{pragmatic-Bayesian method} to account for calibration uncertainty for the \chandra\ effective area within a comprehensive spectral analysis of high-energy spectra. The pragmatic-Bayesian method bases inference for the calibration products only on prior information provided by calibration scientists and their experiments; it ignores information in the current data that might narrow calibration uncertainty. \cite{xu2014fully} enabled a more principled approach, a \textit{fully-Bayesian method}, that dynamically reweights the calibration sample to favor calibration products that are more consistent with the current data and thus allows the data to narrow the calibration uncertainty. Compared with the pragmatic-Bayesian method, the fully-Bayesian method allows the data to simultaneously provide information for estimation of the model parameters and for the calibration product. Along with valid estimates of calibration uncertainty, the parameters can be estimated accurately and efficiently by making full use of information in large-count observed spectra.

In \cite{yu2018incorporating} we developed a Bayesian framework to address the complex problem of conducting spectral analyses that properly account for uncertainty in atomic data. Specifically, we developed methods for propagating uncertainties in \Fexiii\ emissivities to estimate the uncertainties in electron densities using ratios of several EUV emission lines observed by Hinode EIS in a solar active region. The approach was the first to provide an estimate on the uncertainty in each atomic rate, based on the scatter in the values obtained from similar calculations. It relied on a large sample of line emissivities calculated by randomly varying the atomic rates within the bounds of the estimated uncertainties. A similar approach is adopted by \cite{delzanna_etal:2019_n_4} to study uncertainties in derived plasma parameters from \ion{N}{4} lines.  This work takes advantage of a simplification to the model stemming from the fact that the lines considered have the same temperature dependency, so the temperature factor can be ignored in the analysis. The high intensity and high S/N solar data also allowed us to use Gaussian (rather than Poisson) models and to ignore background contamination and contributions from the continuum.  \cite{yu2018incorporating} provides both a pragmatic-Bayesian approach (where the spectral data provide no information for the choice of the atomic physics information) and a fully-Bayesian method where the atomic sample is dynamically reweighted to favor atomic data that are more consistent with the spectral data.

Here, we deploy and compare pragmatic and fully-Bayesian methods to account for uncertainties in atomic data in the context of high-energy spectral analysis, focusing particularly on the accumulated \chandra\ grating data of Capella. The statistical models we develop are significantly more sophisticated than those in \cite{yu2018incorporating} in that they account for the Poisson nature of the X-ray count data, background contamination, and continuum contributions. From a plasma analysis perspective, the model accounts for the effects of atomic data uncertainties in the estimations of both plasma temperature and plasma density, and takes an important step toward a full modeling of the emission measure distribution. We apply our models to ensembles of spectral lines that either depend only on the plasma temperature or on density (with some temperature dependence), and show how we can iteratively update the plasma parameter estimates. Thus, we develop broadly applicable techniques in handling systematic atomic errors and produce a road map for extending them to account for even more complex cases.

We begin, in Section~\ref{sec:astro-models}, by outlining the Capella data and the form our astrophysical models would take if the atomic data were known without errors or uncertainty. Section~\ref{sec:atomic} then describes the nature of uncertainty in the atomic data, how we quantify it, and how we use PCA to avoid sparsity and interpolate within the sample. Our pragmatic and fully-Bayesian statistical framework and our approach to a combined analysis of \Fe\ and \OHe\ spectral lines is laid out in Section~\ref{sec:stats}. Our analyses of the Capella spectra appear in Section~\ref{sec:capella} and the results are discussed in Section~\ref{sec:discuss}. The work is summarized in Section~\ref{sec:summary}.  The details of the optimization algorithm are in Appendix~\ref{sec:app-algs}, and a comparison between the Gibbs and Hamiltonian Monte Carlo methods are in Appendix~\ref{app:compare_algs}.

\section{MODELING CORONAL EMISSION USING FIXED ATOMIC DATA}
\label{sec:astro-models}

We begin with spectral models for a generic ion under the assumption that the necessary atomic data are known without uncertainty. In Section~\ref{sec:atomic} we address uncertainty in the atomic data. Table~\ref{tab:symbol} provides a glossary of the notation used is this article. Where it adds clarity, we place an ``\ionn'' in the superscript of quantities that are ion specific to represent the generic ion. When we discuss particular ions, this superscript is replaced accordingly, for example,  with ``Fe'' for the \Fe \ model and ``O'' for the \OHe \ model.

\begin{table}
\begin{center}
\caption{Glossary of symbols used in the text.}
\begin{tabular}{ m{8.2em}  p{5.0cm} }
 \hline\hline 
\multicolumn{2}{l}{\sl Data and Instruments}\\
Symbol &  Description \\
\hline
\{{\ldots}\} & Generic notation to denote a set of objects \\
$I$ & Oservation index \\
$g$ & Grating index ($h$ for HETG, $l$ for LETG) \\
$o$ & Order ($m$ for $-$ve, $p$ for $+$ve) \\
$\dataso$ & Observed source counts \\
$\databg$ & Observed background counts \\
$\Data$ & Full data, $\Data=(\dataso,\databg)$\\
$\wvl$ & Observed wavelength channel  \\
$\mathcal{W}$ & Set of wavelength channels \\
$\iH$ & Number of wavelength channels \\
$\ratio$ & ratio of background to source collecting area \\
$\exptim$ & Exposure time, $[\text{s}]$ or $[\text{ks}]$ \\
$R$ & Line response function (LRF)  \\
$A$ & Instrument effective area, $[\text{cm}^2]$ or $[\text{cm}^2\text{s}]$ or $[\text{cm}^2\text{ks}]$ \\
\hline\hline
\multicolumn{2}{l}{\sl Atomic} \\
Symbol &  Description \\
\hline
$\ec$ & Emissivity, $[\text{erg cm}^3\text{s}^{-1}]$ \\
$\varepsilon$ & Transformed emissivity,  
$\varepsilon^{\rm Fe} =\ln(\ec^{\rm Fe})$ and
$\varepsilon^{\rm O} =\ln(\ec^{\rm O})$\\
$\emiss$ &
Collections of emissivity realizations\\
$\eigvec$ & Principal components (PCs) of $\emiss$ \\
$\eigvalsqrt$ & Eigenvalues of PCs, representing scaling of each component \\
$\rv$ & Multivariate standard Gaussian variable, 
used to generate $\emiss$ from PCs \\
$(\cdot)^X$ & Superscript represents ion, 
O or Fe \\[5pt]
\hline\hline
\multicolumn{2}{l}{\sl Modeling} \\
Symbol &  Description \\
\hline
$\distas{\rm{indep}}$ & Indicates sampling that is independent and follows the specified distribution. \\
$\log():=\log_{10}()$ & Unless otherwise specified, $\log$ is assumed to be base $10$\\
$\ln()$ & Natural log \\
$\Ne,\log\Ne$  & Plasma electron density, $[\text{cm}^{-3}]$, $\log_{10}[\text{cm}^{-3}]$ \\
$\volume$ & Plasma volume, $[\text{cm}^{3}]$ \\  
$\Te,\logtemp$ & Plasma electron temperature, $[\text{K}]$, $\log_{10}[\text{K}]$  \\
$\param$ & Model parameters, $\param=(\paramso,\parambg)$ \\
$\paramso$ & Plasma parameters, $\paramso^{Fe} = (\volume, \logtemp)$ and $\paramso^{O} = (\log\Ne, \volume, \logtemp)$ \\ 
$\parambg$ & Background parameters \\
$\fx$ & Spectral-line flux at the telescope, $[\text{erg cm}^{-2}\text{s}^{-1}]$ or $[\text{ph cm}^{-2}\text{s}^{-1}]$ \\ 
$\estso$ & Expected line count, [count] \\
$\estcont$ & Expected continuum count, [count] \\
$\ww$ & (True) wavelength, [\AA] \\
$\Omega = \{\ww_l\}$ & Set of (true) wavelengths, [\AA] \\
\hline\hline
\end{tabular}
\label{tab:symbol}
\end{center}
\end{table}

\subsection{\chandra\ data}
\label{sec:chandra_data}

Capella has been observed with several gratings\footnote{There are two grating spectrometers on board \chandra: the HETGS (High-Energy Transmission Grating Spectrometer) and the LETGS (Low-Energy Transmission Grating Spectrometer).} and detector\footnote{Capella data has been obtained using the HETGS and ACIS-S (Advanced CCD Imaging Spectrometer spectrosocpic array) and the HRC-I (High Resolution Camera imaging array); and using LETGS with the ACIS-S and HRC-S (High Resolution Camera spectroscopic array) combinations.  In this study we consider only the ACIS-S datasets.} combinations on board \chandra\ during the course of its mission. Here, we limit our analyses to the ACIS detectors. Together with the $+$ve and $-$ve order dispersion spectra, these yield four distinct datasets (see Table~\ref{tab:chandra_capella}). The datasets are accumulated over several years of individual observations \citep[see][]{2021AJ....162..254M}. Here we primarily use the HETGS+ACIS-S dataset, specifically the MEG $+$ve order ({\tt amp}) to illustrate our analysis. Where available, data from the other datasets\footnote{The HEG does not have significant effective area over the wavelength range of interest relevant to O ions (see Sec~\ref{sec:lineselect}). In fact its sensitivity has decreased with time due to the build up of contamination on ACIS-S, and thus can be used only for the first stage Fe analysis.} are used to either validate the results or describe the magnitude of systematic calibration uncertainty.

\begin{table*}[]
    \centering
    \caption{\chandra\ datasets$^\dag$ of Capella}
    \begin{tabular}{lccccc}
    \hline\hline
    Grating+Instrument & Accumulated & Grating & Data Set & Wavelength & BACKSCAL$^\ddag$\\
    \hfil & Exposure [ks] & Arm & \hfil & Coverage [\AA] &  \\
    \hline
    HETGS+ACIS-S & 622.5 &  MEG$+$1 & \acismegpos &$\approx$1-35 & 1 \\
    \hfil & \hfil & MEG$-$1 & \acismegneg & \hfil & \hfil \\
    \hfil & \hfil & HEG$+$1 & \acishegpos & $\approx$1-18 & 1 \\
    \hfil & \hfil & HEG$-$1 & \acishegneg & \hfil & \hfil \\
    \hline
    LETGS+ACIS-S & 140.9 & LEG$+$1 & \acislegpos & $\approx$1-45 & 15.625 \\
    \hfil & \hfil & LEG$-$1 & \acislegneg & $\approx$1-45 & \hfil \\
    \hline
    \multicolumn{6}{l}{$\dag:$ The specific observations used here are listed at \url{https://doi.org/10.25574/cdc.228}} \\
    \multicolumn{6}{l}{$\ddag:$ Ratio of the areas in which background and source counts are collected.}
    \end{tabular}
    \label{tab:chandra_capella}
\end{table*}

For all observations, we download the event files from the \chandra\ archive \citep{2002SPIE.4844..172R} and reprocess them using the tools in CIAO \citep{2006SPIE.6270E..1VF} to obtain the dispersion wavelengths for each event, and extract the source and background events within the default regions suggested for each grating+detector combination. These events are then binned into histograms with bin widths of $\Delta{w}=0.005$~\AA. We also compute the effective area at each wavelength and combine them, weighted by the exposure times of each observation (see Tables~\ref{tab:fe17_adjusted_lrf} and \ref{tab:ox_adjusted_lrf}).

\subsection{Ion selection and atomic models}
\label{sec:lineselect}

Emission measure techniques applied to Capella have 
indicated that the bulk of emission is narrowly peaked around 6 MK \citep[see, e.g.,][]{brickhouse2000coronal,2003A&A...404.1033A,desai_etal:2005}. 
Such estimates have been obtained with different atomic data but have relied on the common assumption of thermal electron distributions and ionization equilibrium.
Therefore we simplify the problem by approximating the coronal plasma as being isothermal and iso-density, and considering a targeted set of lines that are well-suited to estimate the plasma temperature and density.  We consider only the  strongest lines from the two low-temperature ions, \ion{Fe}{17} and \ion{O}{7}.
We note that some discrepancies in the  intensities of higher-temperature lines of \ion{Fe}{18} and \ion{Fe}{19} in the soft X-rays (around 100~\AA) are present, as discussed in \citep[][]{desai_etal:2005,traebert_beiersdorfer:2021}. Such discrepancies could be due to a complex temperature structure of the hotter plasma, and are not discussed here. 

\ion{Fe}{17} produces the strongest emission lines in the X-rays. For astrophysical plasma densities, this ion does not have metastable levels, i.e., the line intensities have strictly the same density dependence. So the analysis we perform here is similar to that we carried out in \citet{yu2018incorporating}, as the line intensities, except from a scaling factor, only depend on one plasma parameter (the temperature instead of the density in this case).
\cite{delzanna:2011_fe_17} uses solar spectra to show for the first time that accurate temperatures can be obtained from these \ion{Fe}{17} X-ray lines, using the more recent atomic rates also adopted here (previously, factors of 2 discrepancies among the strongest lines were present). 

We stress that the measurement of the electron temperature using the \ion{Fe}{17} lines is independent of the assumption of ionization equilibrium. The only assumption is that electrons have thermal (Maxwellian) distributions. A very strong departure from the thermal distribution would need to be present to significantly affect the line intensities. Some departures are likely present, but have not been observed in the X-rays even in large solar flares. If large departures were present, they would mostly affect the ionization and recombination rates, and shift the formation temperature of the ions, while the effects on the relative intensities of the \ion{Fe}{17} lines would be smaller, see e.g., examples in \cite{dudik_etal:2019} and references therein

To measure the electron densities for \chandra\ MEG/HEG data the best choice are the He-like Ne and O triplets, although in collisionally-dominated plasmas such as those of Capella, the line intensities depend on both density and temperature, see e.g., the review by \cite{delzanna_mason:2018}.

The neon lines are strongly blended with high-temperature lines from several ions in the Capella spectrum \citep[see, e.g.,][]{ness_etal:2003}, and are only sensitive to high densities, so we have opted for the widely used \ion{O}{7} X-ray lines to obtain an estimate of the density. As the line emissivities depend on both electron density and temperature, in this case the analysis is more complex.

Our analysis is carried out in two stages, with the first  focusing on estimating the coronal temperature using density-{\sl insensitive} lines from \ion{Fe}{17} \citep[][see list in Table~\ref{tab:fe17_adjusted_lrf}]{delzanna2011benchmarking}, and the second using density-{\sl sensitive} lines of \ion{O}{7} (see list in Table~\ref{tab:ox_adjusted_lrf}) to estimate both the plasma temperature and plasma density.  The temperature estimate in the second stage is strongly constrained by the temperature in the first stage; the posterior density from the analysis of \ion{Fe}{17} lines is used as an informative Bayesian prior for the analysis of \ion{O}{7} lines.

\subsection{Notation} \label{sec:notation}

{Observations are obtained intermittently, with a separate detector and grating combination for each epoch ($I$). With the HETG, we obtain two independent and simultaneous grating arm observations ($g$=HEG and $g$=MEG) and one with LETG ($g$=LEG). Each grating arm also generates two independent and simultaneous order pairs (conventionally represented as $o$=$+1$ and $o$=$-$1). Each observation comprises a background-contaminated source count $\dataso_{Igo}(\wvl)$ and a pure background count $\databg_{Igo}(\wvl)$ in each of several detector channels $\wvl$ that correspond to the \textit{recorded} photon wavelengths. These cover distinct regions in wavelength, with $\wvl^{Fe}\in[11.1,17.2]$ and $\wvl^{O}\in[17.3,22.2]$, corresponding to where the lines from each of the ions $\ionn\in\{Fe,O\}$ are present.}
Summing over the available observations we obtain $\dataso_{go}(\wvl) = \sum_{I} \dataso_{Igo}(\wvl)$ and
$\databg_{go}(\wvl) = \sum_{I} \databg_{Igo}(\wvl)$. 
{For each of the $L^\ionn \in \{L^{Fe},L^O\}$ spectral lines of interest, we typically select a set of channels that straddle the nominal location of the line to account for the line spread function (LSF).}
We denote the resulting set of $\iH^\ionn$ observed channels wavelengths for each subset by 
\begin{equation}
\mathcal{W}^\ionn = \{\wvl^\ionn_\ih, \ih=1, \ldots, \iH^\ionn\}
\label{eqn:wgrid}
\end{equation}
and the full data set by 
\begin{equation}
\Data^\ionn_{go} = \{(\dataso^\ionn_{go}(\wvl), \databg^\ionn_{go}(\wvl)), \wvl \in \mathcal{W}^\ionn\} \,.
\label{eqn:dataset}
\end{equation}
{Let $\param^\ionn = (\paramso^\ionn, \parambg^\ionn)$ describe the model parameters for ion $\ionn$ and grating-order pair $go$, where $\parambg^\ionn$ represents the background parameter (assumed constant for each ion-grating-order ($\ionn,g,o$) combination), $\paramso^\ionn = (\Ne_{go}, V_{go}^\ionn, \Te_{go}^\ionn)$ represents the plasma parameters: election density, $\Ne$, volume, $V$, and electron plasma temperature, $\Te$. The volume and temperature are tagged here by the ion $\ionn$ because they are computed separately for the corresponding lines $L^\ionn$ ($\Ne$ is estimated only for the lines $L^O$).}
We include $go$ in the subscript of $\paramso$ because we analyze each grating-order pair separately and compare the results. For numerical stability, {descriptive simplicity, and future compatibility (when the emission measure would be parameterized by several shape parameters; 
see Section~\ref{sec:DEMtoflux} below)} we often use the alternative parameterization for the source parameters,
\begin{align}
    \paramso^{\ionn} &= (
    \log \Ne_{go}, V_{go}^{\ionn}, \logtemp_{go}^{\ionn}  )
    \,.
    \label{eq:paramso}
\end{align}

\subsection{DEM, Emissivity, and Source Flux}\label{sec:DEMtoflux}

The flux emitted by the source can be separated into two components, a temperature distribution of emission measures (DEM) that is reliant on the properties of the emitting material (plasma density and volume occupied by plasma at different temperatures), and an emissivity function ($\ec(\ww; \Te, \Ne)$) that arises from atomic physics.  The emissivity describes the power emitted by a unit volume of plasma at a given temperature, $\Te$ (typically between $\approx$10$^{6-7}$~K, and density, $\Ne$ (typically $\ll$10$^{14}$~cm$^{-3}$), at a wavelength $\ww$ from both line transitions and continuum emission.  This emission is optically thin, i.e., emitted photons escape without scattering and absorption. 

The DEM allows us to describe the physical conditions in the corona, specifically through a measure of the volume $\volume$ of plasma that exists at a given density and temperature,
\begin{equation}
{\rm DEM}(\Ne,\Te,\volume) = \Ne^2(\Te)~\frac{dV(\Ne,\Te)}{d\logtemp} \,.
\label{eq:DEM}
\end{equation}
In most analyses of stellar coronae, an isodensity plasma is assumed, with $V(\Ne,\Te)=V(\Te)$. In this work, we further assume an isothermal plasma, which is a reasonable approximation in the case of Capella (see Section~\ref{sec:intro}), with 
\begin{equation}
    \textrm{plasma~volume}=V\cdot\delta(\logtemp_\textrm{iso}-\logtemp) \,,
    \label{eq:volume}
\end{equation}
where $\delta(x-x_0)$ is the Dirac delta function, with value $1$ if $x = x_0$ and $0$ otherwise, fixing the plasma temperature at $\Te_{\rm iso}$ and the total emission measure at
\begin{equation}
    {EM} = \int {\rm DEM(\Ne,\Te,V)}~d\logtemp = \Ne^2\volume \,.
    \label{eq:EM}
\end{equation}
Thus, different values of $\logtemp$ describe coronae of different temperatures\footnote{In general, the shape of the DEM can be described as a continuous or discrete function with many parameters.  For example, a coronal loop can be characterized as a power-law in $\Te$, with 3 parameters, as DEM$(\Te;\alpha_1,\alpha_2,\alpha_3) = \alpha_1\cdot\Te^{\alpha_2}$ for $\Te<\alpha_3$ (notice that one more parameter is needed than in the case of the isothermal plasma).  Other forms, such as multiple isothermal components, log-normals, mixtures of log-normals, piecewise power-laws, Chebyshev polynomials, etc., may be used.  Additional parameterization to describe variations in $\Ne$ are also possible, though as a practical matter for lines that are not density sensitive, $\Ne^2$ and $\volume$ are degenerate in total emission measure $EM$.  In any case, the parameterization must be sufficiently constrained such that the number of estimated parameters does not exceed the number of data points.  In the discussion in Section~\ref{sec:atomic}, we also include PCA parameters to describe the uncertainties in the atomic emissivities, and they must also be included in the total count of parameters.  This is one of the reasons we limit our analysis to an isothermal and isodensity plasma.  The lines list must be expanded if a more complex DEM shape is to be modeled.}, which consequently yield (see below) different model spectra $\estso_{Igo}(\wvl;\paramso)$ and $\estcont_{Igo}(\wvl;\paramso)$ (see Equations~\ref{eq:estso} and \ref{eq:estcont}).

Combining the emissivity and the DEM, the flux from a line $l$ at intrinsic wavelength $\ww_l$, incident at the telescope from a source at distance $d_*$ can be written (using Equations~\ref{eq:DEM} to \ref{eq:EM}) as
\begin{eqnarray}
\fx(\ww_l;\paramso) &=& \frac{1}{4 \pi d_*^2}\int         \ec(\ww_l;\Ne_{go},\Te_{go})~{\rm DEM}(\Ne,\Te,\volume)~d\logtemp \nonumber \\
               &=& \frac{\Ne_{go}^2}{4 \pi d_*^2} \int \ec(\ww_l;\Ne,\Te)~\frac{dV(\Te)}{d\logtemp}~d\logtemp \nonumber \\
               &=& \frac{\Ne_{go}^2~\cdot~\volume_{go}}{4 \pi d_*^2} ~ \ec(\ww_l;\Ne_{go}, \logtemp_{{\rm iso},{go}}) \,,
\label{eq:estfx_delta}
\end{eqnarray}
where $\paramso = (\log\Ne_{go}, \log\volume_{go}, \logtemp_{{\rm iso},go})$.
Similarly, the continuum emission can  be computed as
\begin{eqnarray}
    c(\ww;\paramso) &=& \frac{\Ne_{go}^2~\cdot~\volume_{go}}{4 \pi d_*^2} \times \nonumber \\
    &&\left[ \ec_{ff} + \ec_{fb} + \ec_{pc}\right](\ww;\Ne_{go},\logtemp_{{\rm iso},{go}}) \,,
\end{eqnarray}
where $\ec_{ff}, \ec_{fb}$, and  $\ec_{pc}$ represent the free-free (Bremsstrahlung), free-bound (radiative recombination), and psuedo-continuum (arising from unresolved weak lines) components. Notice that we explicitly include the subscripts denoting grating and order to emphasize that the parameter estimates are obtained for specific datasets.  Going forward, for the sake of simplicity, we shall drop the subscripts and include them only if there is ambiguity, and further use $\logtemp$ to represent $\logtemp_{\rm iso}$.

\subsection{The Basic Model}
\label{sec:basic_model}

{We model the counts $\dataso_{go}(\wvl)$ as a combination of the expected counts due to individual spectral lines, the continuum, and the background.}
First, the \textit{expected spectral-line count} in channel $\wvl$ under a spectral model with parameter $\paramso$ is given by
\begin{equation}
\estso_{Igo}(\wvl^{\ionn};\paramso^{\ionn}) = \sum_{\il=1}^{L^{\ionn}} \lrflion \cdot \arfl \cdot \fx(\ww_\il;\paramso^{\ionn}) \cdot \exptim_{I},
\label{eq:estso} 
\end{equation}
where the sum is over the wavelengths of the $L^{\ionn}$ analyzed spectral lines, i.e., $\Omega^{\ionn} = \{\ww_\il, \il=1, \cdots, \iL^{\ionn}\}$,
$\lrfl$ is the Line Response Function (LRF) of the instrument with grating-order pair $go$, $\arfl$ is the effective area of the instrument with observation $I$ and grating-order pair $go$,
$\fx(\ww_\il;\paramso^{\ionn})$ is the spectral-line flux \textit{at the telescope} at (true) wavelength $\ww_\il$, and $\exptim_{I}$ is the exposure time of observation $I$. 
For computational efficiency, we use an analytical LRF appropriate for \chandra\ gratings, specifically\footnote{This choice of LRF is a better representation of the point spread function (PSF), integrated over the cross-dispersion direction, see Equation~9.1 of the \chandra\ Observatory Proposers' Guide, \url{https://cxc.harvard.edu/proposer/POG/html/chap9.html\#tth_sEc9.3.3}.  This functional form conveniently allows us to adjust for distortions due to detector non-uniformity or contamination by weak lines, by allowing the width $\sigma$ to be adjusted to match the observed widths.  The value of $\nu=4$ is a useful intermediate functional form between a heavy-tailed Lorentzian ($\nu=1$) and a broad core Gaussian ($\nu\gg1$) profiles, and is held fixed here to avoid making substantial changes to the adopted calibration.}
\begin{equation}
\lrf = \frac{\Gamma(\frac{\nu+1}{2})}{\sigma \sqrt{\nu \pi} \Gamma(\frac{\nu}{2})} \left[1 + \frac{(\wvl-\ww)^2}{\nu \sigma^2}\right]^{-\frac{\nu+1}{2}}, \label{eq:lrf}
\end{equation}
which is the probability density function of a generalized $t$-distribution with degrees of freedom $\nu \equiv 4$, location parameter $\ww$, and scale parameter $\sigma$, which is $\approx$0.0115 for the {\tt amp} dataset\footnote{See Table~8.1 of the \chandra\ Observatory Proposers' Guide, \url{https://cxc.cfa.harvard.edu/proposer/POG/html/chap8.html\#tth_tAb8.1}}.  Because the instrument is not perfect, the particular values of $\ww$ and $\sigma$ can differ for different spectral lines.

The \textit{expected continuum count} in channel $\wvl$ is
\begin{multline}
\estcont_{Igo}(\wvl^{\ionn}; \paramso^{\ionn}) = \\
\sum_{\ih=1}^{\iH}  \delta(\wvl^{\ionn}-\ww_\ih) \cdot \arfh  \cdot \ctnm(\ww_\ih, \paramso^{\ionn}) \cdot \exptim_{I},
\label{eq:estcont}
\end{multline}
where $\{\omega_\ih\}$ is a binning of the true wavelenghts that coincides with $\cal W$, 
and $\ctnm(\ww_\ih,\Te)$ is the continuum flux at the telescope at wavelength $\ww_
\ih$ and temperature $\Te$. 
Note that since the continuum is smooth over the widths of the LRF, especially around the lines of interest, we simplify our numerical estimation by ignoring the LRF for the continuum in Equation~\ref{eq:estcont}.

Thus, for each grating-order pair, we model the observed counts as
\begin{align}
      \dataso_{go}(\wvl^{\ionn}) \mid \paramso^{\ionn}, \ \parambg^{\ionn} & \distas{\rm{indep}} \operatorname{Poisson} \big( \estso_{go}(\wvl^{\ionn};\paramso^{\ionn}) + \nonumber  \\  
      & \ \ \ \ \ \ \ \ \ \ \ \ \  \estcont_{go}(\wvl^{\ionn};\paramso^{\ionn}) + \parambg^{\ionn} \big), \label{eq:likeli_so_pois} \\
      \databg_{go}(\wvl^{\ionn}) \mid \parambg^{\ionn} & \distas{\rm{indep}} \operatorname{Poisson} \big( \ratio_{go} \cdot \parambg^{\ionn} \big), \label{eq:likeli_bg_pois} 
\end{align} 
where $\distas{\rm{indep}}$ indicates independent observation from a given distribution, $\estso_{go}(\wvl^{\ionn};\paramso^{\ionn})= \sum_{I} \estso_{Igo}(\wvl^{\ionn};\paramso^{\ionn})$, $\estcont_{go}(\wvl^{\ionn}; \paramso^{\ionn}) = \sum_{I} \estcont_{Igo}(\wvl^{\ionn}; \paramso^{\ionn})$, $\parambg^{\ionn}$ represents the \textit{expected background count}, and $\ratio_{go}$ is
the ratio of background and source collection areas, which are each computed as the sums over the observations of the product of the exposure times and the geometric areas over which the respective counts are collected.

The modeling is carried out separately for each subset $\{\wvl^\ionn\}$ of ions, 
and for each grating-order combination ($g,o$).
For simplicity, we assume that the shape of the expected continuum count is fixed, i.e., $\estcont_{go}(\wvl; \paramso)$ is known in advance for each channel $\wvl$ at a given $\paramso$
and that the expected background count is the same across channels for 
{a given $\{\ionn,g,o\}$ combination.}

Because we conduct analyses separately for each ion-grating-order combination, from here on we simplify notation by suppressing the superscript $\ionn$ and subscript $go$ except where specific clarification is required. 

\subsection{Preprocessing the spectral data} \label{sec:preprocessing}

We preprocess the full \chandra\ dataset in order to focus the analysis on the lines of interest and to minimize the impact of systematic effects like contamination from nearby lines and atomic and calibration uncertainties on line locations and widths.

First, we filter the data to include only those channels nearest to the nominal locations of the lines of interest.  The typical width used is $\pm$4 channels, corresponding to $\pm$0.02~\AA\ for ACIS spectra.
Furthermore, we omit lines and channels where contamination from a nearby line could affect the estimated flux; for instance, we omit the 16.004~\AA\ Fe~line because it is weak and contaminated by a nearby \OHe\ line \citep{delzanna2011benchmarking}.
This results in a collection of non-contiguous ranges of channels $\mathcal{W}$ (see Equation~\ref{eqn:wgrid}), with $\iH=100$ channels used in the \Fe\ analysis and $\iH=112$ channels used in the \OHe\ analysis.

\begin{figure}[!htbp]
\centering
\includegraphics[width=\linewidth]{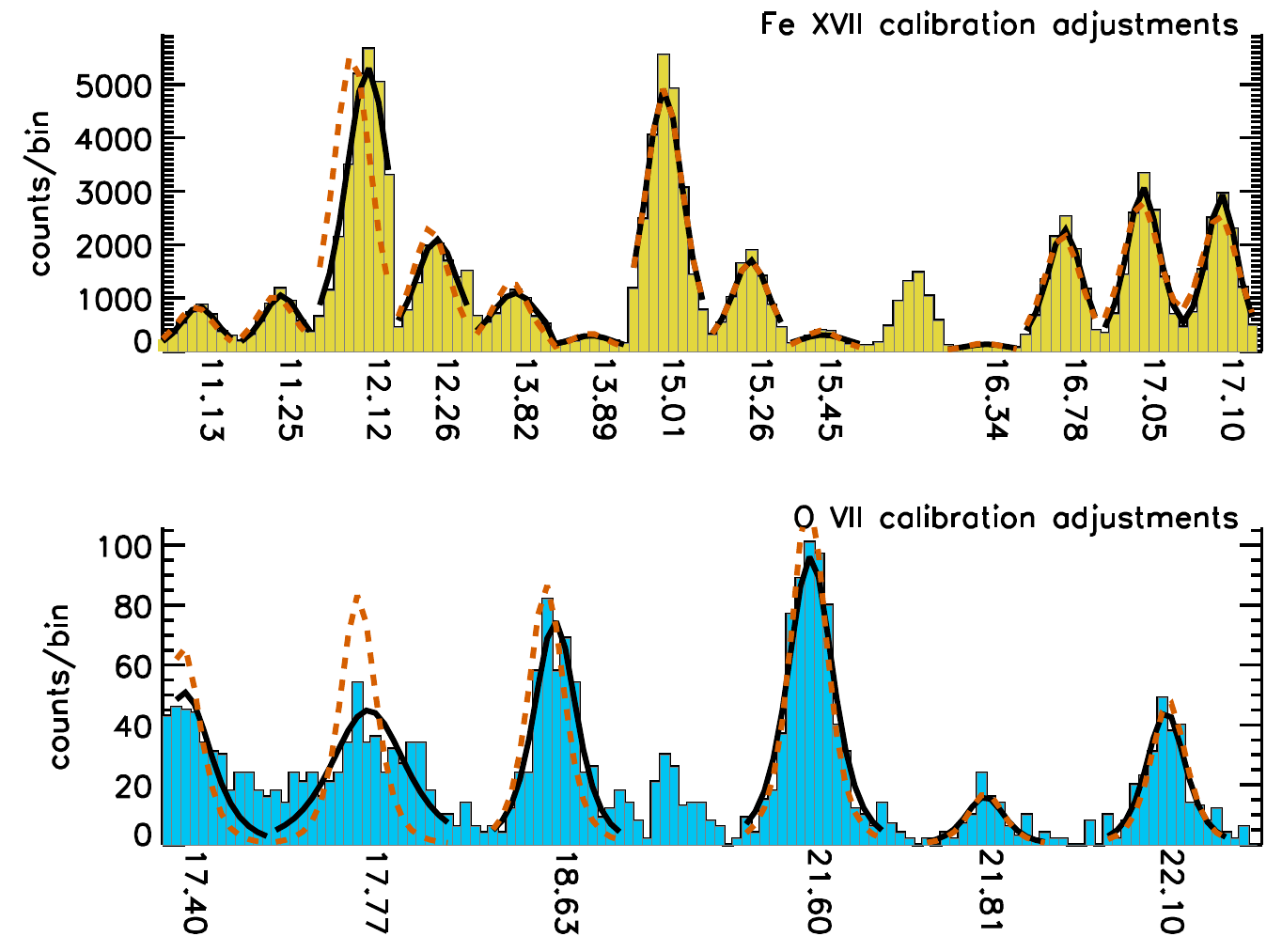}
\caption{
Describing the calibration alignment adjustments for the \Fe\ ({\sl top panel}) and \OHe\ ({\sl bottom panel}). The shaded histograms depict the observed {\tt amp} counts spectrum (Table~\ref{tab:chandra_capella}) over wavelength bins (of size 0.005~\AA) in the lines of interest.  The gaps between lines are excluded for the sake of visibility, and the lines of interest are marked.  The dashed red curves depict the nominal LRF, and the solid black curves show the LRF adjusted to remove wavelength and width misalignments (see text).  The heights of each LRF are arbitrarily scaled to match the observed intensity in each line.
}
\label{fig:fe17_spectrum_shift_process}
\end{figure}

Second, because the detectors are not perfectly calibrated, we may observe small mismatches ($\lesssim$5~m\AA, which corresponds to the native pixelization on ACIS) between the nominal and estimated locations of the lines of interest, as well as between the estimated line widths and those expected from the instrument LRF.  We carry out fits to the spectra by adjusting the LRF separately for each line in each grating.  We accomplish this by first constructing a predicted count spectrum conditional on a fixed $\logtemp$ and $\Ne$ ($\logtemp = 6.739$ for \Fe, $\logtemp = 6.571$ and $\log \Ne = 9.310$ for \OHe) in order to ensure that the lines in the predicted spectrum are correct identifications and to minimize wavelength drifts due to contaminating lines that may appear at different temperatures and densities.  We obtain minimum $\chi^2$ estimates of the location and scale parameters, $\omega$ and $\sigma$, of the LRF given in Equation~\ref{eq:lrf} for each line-grating combination. 
The fitted values for \Fe\ and \OHe\ {\tt amp} spectral lines appear in Tables~\ref{tab:fe17_adjusted_lrf} and \ref{tab:ox_adjusted_lrf}, respectively. The fitted values for other line-grating combinations appear in Appendix~\ref{app:LRF_other_data_sets}.
The results for the LRF-adjusted {\tt amp} \Fe\ and \OHe\ spectral data appear as solid black lines in Figure~\ref{fig:fe17_spectrum_shift_process}, illustrating a much improved overall fit.  Notice that the counts around the 16.004~\AA\ line, which has been ignored in the analysis, are shown in Figure~\ref{fig:fe17_spectrum_shift_process}, but an improved LRF fit has not been carried out for it.

\begin{table}
\begin{center}
\caption{A summary of the best-fit scale, $\hat{\sigma}$, best-fit location, $\hat{\ww}$, and nominal location, $\ww$, of LRF for each \Fe \ {\tt amp} spectral line.}
\begin{tabular}{c c c c } 
 \hline
 \hline
$\ww \ (\mathring{\text{A}})$ & $\hat{\ww} \ (\mathring{\text{A}})$ &  $\hat{\sigma}$  & Effective Area $(\text{cm}^2 k\text{sec})$  \\
 \hline
$11.129$ & $11.131$ & $0.0090$ & $14560$  \\
$11.250$ & $11.253$ & $0.0085$ & $13640$  \\
$12.124$ & $12.122$ & $0.0105$ & $10570$  \\
$12.264$ & $12.267$ & $0.0105$ & $10010$  \\
$13.825$ & $13.827$ & $0.0115$ & $6024$  \\
$13.890$ & $13.891$ & $0.0110$ & $5894$ \\
$15.013$ & $15.013$ & $0.0090$ & $3335$  \\
$15.262$ & $15.262$ & $0.0090$ & $2866$  \\
$15.453$ & $15.454$ & $0.0125$ & $2826$  \\
$16.336$ & $16.337$ & $0.0105$ & $2395$  \\
$16.776$ & $16.777$ & $0.0085$ & $2094$  \\
$17.051$ & $17.048$ & $0.0080$ & $1918$  \\
$17.096$ & $17.092$ & $0.0075$ & $1908$  \\
\hline\hline
\end{tabular}
\label{tab:fe17_adjusted_lrf}
\end{center}
\end{table}

\begin{table}
\begin{center}
\caption{A summary of the best-fit scale $\hat{\sigma}$, best-fit location, $\hat{\ww}$, and nominal location, $\ww$, of LRF for each \OHe \ {\tt amp} spectral line.}
\begin{tabular}{c c c c } 
 \hline
 \hline
$\ww \ (\mathring{\text{A}})$ & $\hat{\ww} \ (\mathring{\text{A}})$ &  $\hat{\sigma}$  & effective area $(\text{cm}^2 k\text{sec})$  \\
 \hline
$17.396$ & $17.3964$ & $0.0123$ & $1692$  \\
$17.768$ & $17.7724$ & $0.0148$ & $1461$  \\
$18.627$ & $18.6319$ & $0.0103$ & $1130$  \\
$21.602$ & $21.6030$ & $0.0112$ & $313.6$  \\
$21.805$ & $21.8044$ & $0.0056$ & $252.3$  \\
$22.101$ & $22.0995$ & $0.0073$ & $182.4$  \\
 \hline
 \hline
\end{tabular}
\label{tab:ox_adjusted_lrf}
\end{center}
\end{table}

\section{Uncertainties in the Atomic Data}
\label{sec:atomic}

The models in Section~\ref{sec:astro-models} assume that the emissivities are known without uncertainty. Here we describe our preliminary analyses to quantify their uncertainty. The result is a collection of $M$ realizations of the \textit{atomic line emissivities} that represent the population of possible curves. We denote the collection by $\mathcal{M}^{X} = \{ \ec^{X}_{(\rvm)}(\ww; \Ne, \Te) , \ww \in \Omega, m = 1,\ldots,M \}$,  where $\rvm$ indexes of the emissivity realizations.  From a statistical perspective, we view $\mathcal{M}$ as a sample from the prior distribution of the emissivities, see Sections~\ref{sec:atomic_pca} and \ref{sec:stats}.

\subsection{Uncertainties in the atomic data for \Fe}
\label{sec:atomic_fe}

Following the \ion{Fe}{13} study of \citet{xu2014fully}, we compare the atomic rates calculations of \cite{loch_etal:06} and \cite{liang_badnell:2010_ne-like} to provide an estimate of the uncertainty for each individual rate. Ideally, one would use more than two calculations to estimate uncertainties in atomic rates. However, only these two scattering calculations are of a similar level of sophistication. They both adopt the close-coupling $R$-matrix suites of codes (see the UK APAP network\footnote{www.apap-network.org}  website),  originally developed for the Iron Project. \cite{loch_etal:06} include 139 states (up to $n$=4) in their close-coupling expansion (CCE). The orbitals are determined from a Dirac-Fock structure calculation using the GRASP program, developed by I. Grant and collaborators (see e.g., \citealt{grasp92}). \cite{liang_badnell:2010_ne-like} employ a larger target, with a CCE including 209 states (up to $n$=5). Their R-matrix calculation adopts the ICFT approximation (see a review by \citealt{badnell_etal:2016}), while the target orbitals are determined using the {\sc autostructure} program \citep{badnell:2011}. We note that the level ordering of the two calculations is not the same. 

\cite{liang_badnell:2010_ne-like} find that their main atomic rates are similar to those of \cite{loch_etal:06}. Indeed \cite{delzanna:2011_fe_17} shows that the line emissivities obtained from these two sets of rates are very similar, and uses them to show for the first time that the \ion{Fe}{17} lines can be used to measure the electron temperature. Using the best solar observations of the X-ray lines, \cite{delzanna:2011_fe_17} finds very good agreement (within 10\%) between predicted and measured line intensities for an isothermal plasma.  The long-standing discrepancies that have been present when distorted wave (DW) calculations were adopted were finally resolved, although we note that problems in modelling other plasma are still being highlighted in the literature \citep[see, e.g.,][]{2020PhRvL.124v5001K}. The main differences with the DW calculations are caused by resonances that significantly increase the collision strengths of the 2p--3s transitions, the same problem that was resolved for the  \ion{Fe}{18} lines \citep{delzanna:2006_fe_18}.

The emissivity of a spectral line is proportional to the population of the upper level and the spontaneous transition probability (i.e., the radiative rate).  \Fe\  does not have metastable levels, which implies that the intensities of the lines all have the same dependence on the electron density. Moreover, its states are mostly populated only by collisional excitation (CE) from the ground level,  2s$^2$ 2p$^6$ $^1$S$_{0}$. Only the levels No.2 and 4, the 2s$^2$  2p$^5$ 3s $^3$P$_{2,0}$, reach a very small relative population of 10$^{-4}$ at a high density of 10$^{13}$ cm$^{-3}$.  Thus, for completeness and simplicity we built an ion model that includes only the CE from these two states, as well as the ground state. 

A comparison of the thermally-averaged cross-sections calculated by \cite{loch_etal:06} and \cite{liang_badnell:2010_ne-like} at 2 MK shows that most transitions from these three states differ by  between 1 and 20\%, with the weaker ones progressively diverging by large factors. We set a lower limit to the uncertainty of 5\% and an upper limit of 100\%.

Regarding the radiative rates, for most cases differences between the \cite{loch_etal:06} and \cite{liang_badnell:2010_ne-like} values are between 1 and 30\%. For the CE rates, we set a lower limit to the uncertainty of 5\% and an upper limit of 100\%.

We obtain 1000 Monte Carlo calculations of line emissivities, by randomly varying (independently) the two sets of atomic rates (CE and radiative), using a modified version of the CHIANTI\footnote{\url{https://www.chiantidatabase.org}} version 10 \citep{chianti_v10} codes. We use the IDL function {\it randomn} to randomly vary each rate within its estimated uncertainty. We assume a normal distribution, in the sense that 
if, e.g., an uncertainty is 10\%, then 95\% of the simulated values vary within $\pm$ 20\%. We take the atomic rates of \cite{loch_etal:06} as a baseline. 

\subsection{Uncertainties in the Atomic Data for \OHe} \label{sec:atomic_oxygen}

As the calculations are relatively simple to run, we adopt a different approach for \OHe. Specifically, we run a series of ICFT $R$-matrix calculations to estimate the scatter in the atomic rates and hence their uncertainties. We use a target structure from {\sc autostructure}, including in the CCE 49 states up to $n=5$. The Thomas-Amaldi-Fermi scaling parameters for the bound orbitals in the central potential model are set to 1. The  exchange calculation is performed up to $J$=16.5 while the non-exchange calculation is performed up to $J$=35.5, followed by a top-up calculation for higher partial waves. Eighty continuum basis orbitals are used, giving a smallest maximum basis-orbital energy of 298.6757 Rydbergs. Different energy meshes are used for the inner and outer calculations. Dipole and Born limits are used from the last calculated energy points up to infinite energy to calculate the effective collision strengths. This is used as a baseline calculation.

We then vary the scaling parameters randomly to obtain ten further different targets, bound by the hydrgen-like and lithium-like structures.  As the CE and radiative rates are directly related to the target structure, we carry out ten ICFT $R$-matrix calculations, to obtain ten different sets of rates. Despite the significantly large variations in the target structure, differences in the rates and line emissivities are relatively small, of the order of a few percent. To apply a similar procedure as for the \ion{Fe}{17}, we take  an uncertainty estimated from the maximum difference among the 11 calculations for each rate. We set a lower limit for the uncertainty as 2\%, and an upper limit of 80\%, for both the CE (considering the values at 1 MK), and the radiative rates. We generated $10,000$ Monte Carlo calculations of line emissivities, by randomly varying (independently) the two sets of atomic rates,  as in the \ion{Fe}{17} case. As we are interested in the strongest X-ray lines from He-like O, and as they are sensitive to both electron temperatures and densities, we have performed the calculations over a grid of temperatures and densities.

\subsection{PCA-based model for atomic uncertainties} \label{sec:atomic_pca}

We propose a statistical model for the collections of emissivity realizations, $\emissfe$ and $\emissox$, in order to more readily incorporate them into our analyses. Specifically, we propose reduced-dimension multivariate Gaussian distributions. Following \citet{lee2011accounting} we use principal component analysis (PCA) to identify a low-dimensional orthogonal linear transformation of the emissivites that maintains the bulk of the variability exhibited in  $\emissfe$ and $\emissox$. 

To improve the quality of the principal component Gaussian models for $\emissfe$ and $\emissox$ we transform the individual emissivity replicates using a variance stabilizing transform.  Specifically, we compute $\ln\ec$ and carry out a PCA in the transformed emissivity space.\footnote{A variance stabilizing transform maps a function $f(y)$ such that the dependence of the variance on the value of $f(y)$ is reduced.  This can happen when fluctuations are fractional relative to the value, and taking the log makes the scatter the same at all levels.  Other transforms, like $\sqrt{\ec}$, are also apposite.  We choose the log transformation because of the large dynamic range present in the emissivities.} We demonstrate below in Sections~\ref{sec:atomic_pca_fe} and~\ref{sec:atomic_pca_ox} that the adopted $\ln$ transformation works well empirically, i.e., the 68\% and 95\% quantiles of the PC reconstructed emissivities match the quantiles of the original replicate sets $\emissfe$ and $\emissox$ well (see Figures~\ref{fig:loge_pca_fe17} and~\ref{fig:ox_loge_pca}.
We denote the transformed variables by $\varepsilon$ and their means across the realizations in $\emissfe$ and $\emissox$ by
\begin{equation}
\bar{\varepsilon}^{\ionn}= \overline{\ln \ec}^{\ionn} = {1\over M} \sum_{\m=1}^M \ln(\ec_{(\m)}) \ \hbox{ with } \ \ec_{(\m)} \in \emiss^{\ionn}.
\end{equation} 

Similarly, we let $\Sigma^{\rm Fe}$ be a diagonal matrix with diagonal elements equal to the variances across $\emissfe$ of the components of the transformed \Fe\ emissivity. $\Sigma^{\rm O}$ is defined analogously. 

Working on the transformed scales, and before applying PCA, we standardize each variable by subtracting off the mean and dividing by the standard deviation of each variable\footnote{Specifically we apply PCA to the sets 
$$
\Big\{\left(\Sigma^X\right)^{-{1\over 2}} (\varepsilon^X_{(m)} - \bar{\varepsilon}^X),\  \hbox{ for } \ m=1,\ldots, M \Big\}, \text{where } X\in\{\hbox{Fe, O}\}.
$$}. 
With PCA, we are able to reduce the dimension of the (transformed) \Fe\ emissivities from $308$ to $7$ while preserving $55\%$ of the total variance of the standardized variables. For the \OHe\ emissivities, we can reduce the dimension from $7038$ to $6$ while preserving $97\%$ of the variance.  
Mathematically, we represent the Gaussian model for the emissivities in terms of a random replicate, $\varepsilon^{X}_{\text{rep}}$, from the model, where $X\in\{\hbox{Fe, O}\}$, again using $\varepsilon$ to emphasize the transformed scales. Specifically,
\begin{equation}
\varepsilon^{X}_{\text{rep}} (\rv^X) = \bar{\varepsilon}^{X} + \left(\Sigma^{X}\right)^{1/2}\sum_{j=1}^J \rv^{X}_j \ \eigvalsqrt^{X}_j \ \eigvec^{X}_j,
\label{eq:pca_emiss_replicate}
\end{equation}

where $J$ is the reduced-dimension of the Gaussian model  (i.e., $J=7$ for $\emissfe$ and $J=6$ for $\emissox$), $\eigvec^{X}_1, \ldots, \eigvec^{X}_J$ are the first $J$ principal components with corresponding scales $\eigvalsqrt^{X}_1 \geq \eigvalsqrt^{X}_2 \geq \ldots, \geq \eigvalsqrt^{X}_J$, $r^X=(r^X_1, \ldots, r^X_J)$ represents the PCA projections, and we model $r^X$ as a $J$-dimensional mean-zero multivariate Gaussian with variance-covariance matrix equal to $I$, the identity matrix (i.e., unit variances and uncorrelated). In the the spectral analyses described in Section~\ref{sec:stats}, we summarize the Gaussian models for $\emissfe$ and $\emissox$ as prior distributions on $\rvfe$ and  $\rvox$, respectively:
\begin{align}
  \rvfe &\sim \hbox{MV  Gaussian} (0, I) \ \hbox{ and } \label{eq:prior_rvfe} \\
    \rvox &\sim \hbox{MV Gaussian} (0, I). \label{eq:prior_rvox}
\end{align}

In Equation~\ref{eq:pca_emiss_replicate}, $\varepsilon^X_{\text{rep}}(\rv^X)$ is written as a function of $\rv$ to emphasize that under the model we can compute the high-dimensional emissivity as a linear function of the low dimensional $\rv^X$. Because $\rv^X$ follows a multivariate Gaussian distribution, so does its linear function $\varepsilon^{X}_{\text{rep}} (\rv^X)$, albeit a degenerate Gaussian. As described in Sections~\ref{sec:atomic_fe} and \ref{sec:atomic_oxygen}, we obtain $1000$ and $10,000$ replicate emissivity curves for $\emissfe$ and $\emissox$ respectively. We discuss the fit of the reduced-dimension multivariate Gaussian distributions to $\emissfe$ and $\emissox$ in Sections~\ref{sec:atomic_pca_fe} and \ref{sec:atomic_pca_ox}.

As detailed in Section~\ref{sec:stats}, we account for uncertainty in the emissivity curves by treating $\rvfe$ and $\rvox$ as unknown model parameters with prior distributions given in Equations~\ref{eq:prior_rvfe} and \ref{eq:prior_rvox} and that are estimated (with error bars) from the data. They enter the statistical model of Section~\ref{sec:basic_model} via the expected spectral-line counts, $\estso_{Igo}(\wvl;\paramso)$ and thus we include $\rv$ as an argument of $\estso_{Igo}$ where appropriate.

\subsubsection{Gaussian Prior on \Fe \ Emissivity}
\label{sec:atomic_pca_fe}

\begin{figure*}
\centering
\includegraphics[width=\linewidth]{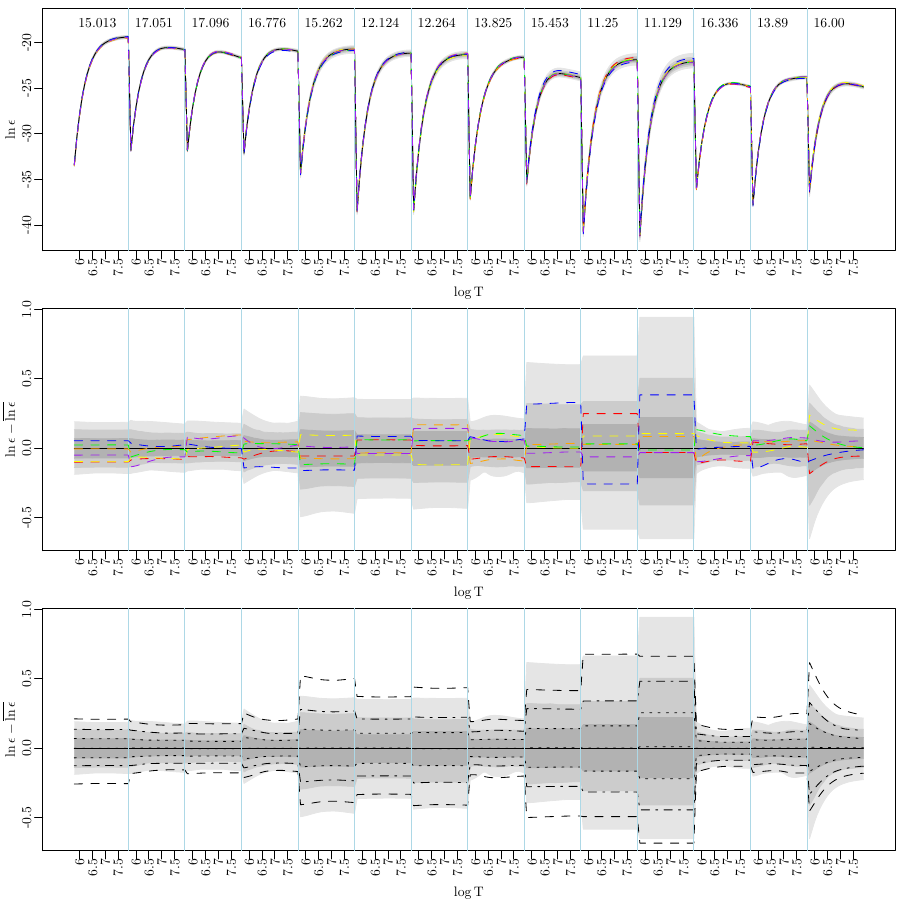}
\caption{Comparing $\emissfe$ with its PCA-based Gaussian model.  Top panel: light, dark, and darker gray areas give point-wise intervals for the full range,  middle $95\%$, and middle $68.3\%$ of $\emissfe$ on the $\ln$ scale. Six randomly selected curves from $\emissfe$ are plotted as colored dashed curves and $\overline{\ln \ec}^{\rm Fe}$ is plotted as a solid black curve. Middle panel: same as top panel, but plotting the residuals,  $\ln{\ec_{(\m)}} - \overline{\ln \ec}^{\rm Fe}$. Bottom Panel: dashed, dotted-dash, and dotted lines compare point-wise intervals for the full range, middle $95\%$, and middle $68.3\%$ of $1,000$ replicate emissivity curves sampled under the PCA-based Gaussian model with the same intervals of $\emissfe$ (shades of grey).}
\label{fig:loge_pca_fe17}
\end{figure*}

Figure~\ref{fig:loge_pca_fe17} illustrates the PCA-based Gaussian model for $\emissfe$ (on the natural log scale). The complex structure of the \Fe \ emissivities is illustrated in Figure~\ref{fig:loge_pca_fe17} (top panel). A random selection of six $\ln{\ec_{(\m)}}$ from $\emissfe$ are plotted as colored dashed lines and compared with the average over the transformed $\emissfe$, i.e., $\overline{\ln \ec}^{\rm Fe}$, plotted as a solid black line. The light, dark, and darker gray areas represent the point-wise full range, middle $95\%$, and middle $68.3\%$ of the $1,000$ transformed emissivity curves in $\emissfe$. The ensemble $\emissfe$ forms a complex tangle that appears to defy any systematic pattern. Figure~\ref{fig:loge_pca_fe17} (middle panel) is the same as the top panel except it plots the residuals, $\ln{\ec_{(\m)}} - \overline{\ln \ec}^{\rm Fe}$, in order to magnify the structure in $\emissfe$. The dashed, dotted-dashed, and dotted lines, in Figure~\ref{fig:loge_pca_fe17} (bottom panel) superimpose the point-wise full range, middle $95\%$, and middle $68.3\%$ intervals of 1000 emissivity curves sampled under the Gaussian model on the corresponding intervals of $\emissfe$ (shades of gray). The correspondence between $\emissfe$ and the PCA-based Gaussian model is quite good, especially for the $68.3\%$ intervals.

\subsubsection{Gaussian Prior on \OHe \ Emissivity}
\label{sec:atomic_pca_ox}

\begin{figure*}
\centering
\includegraphics[width=0.85\linewidth]{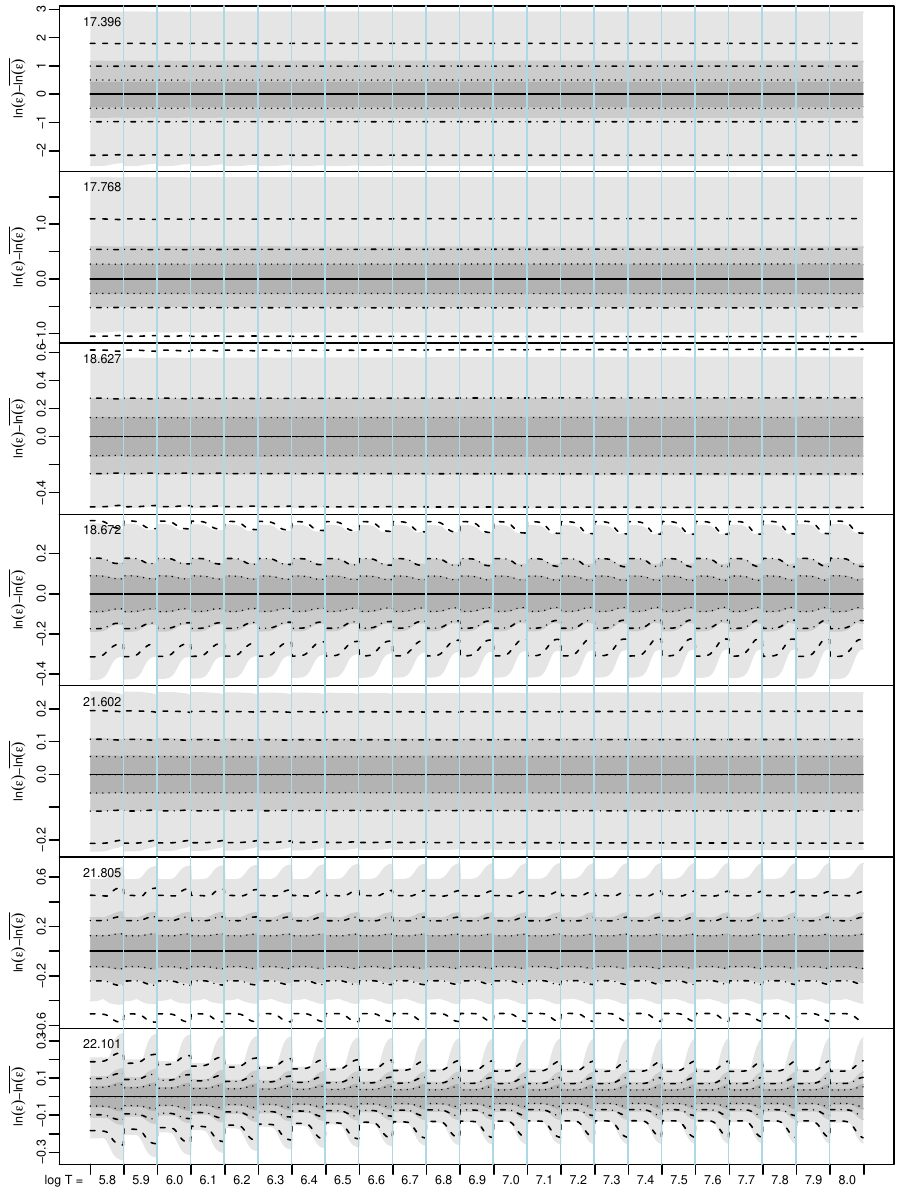}
\caption{Comparing $\emissox$ with its PCA-based Gaussian model. Each panel corresponds to one of the seven \OHe \ spectral lines; the wavelength is given in the upper left of each panel. The mean over $\emissox$ on the $\ln$ scale, $\bar{\varepsilon}^{\rm O}$, is subtracted off to magnify the structure. The zero line is plotted as a solid black curve. The light, dark, and darker gray areas cover the point-wise full range, middle $95\%$, and middle $68.3\%$ of $\emissox$ (again on the $\ln$ scale). The dashed, dotted-dashed, and dotted lines provide the same intervals for $10,000$ replicate emissivity curves generated under the PCA-based Gaussian model. In each panel, the horizontally-arranged blocks correspond to the values of the equally-spaced temperature grid, $\logtemp \in (5.8, 8.0)$. Within each $\logtemp$ block, $\log \Ne$ increases from left to right in the interval $(8.0, 13.0)$.}
\label{fig:ox_loge_pca}
\end{figure*}

Figure~\ref{fig:ox_loge_pca} illustrates the PCA-based Gaussian model for $\mathcal{M}^{O}$ (on the $\ln$ scale). 
Significant compression is achieved because only 6 principal components are needed to capture 97\% of the variability in $\emissox$. Each panel corresponds to one of the \OHe \ spectral lines and plots the residuals $\ln{\ec_{(\m)}} - \overline{\ln \ec}^{\rm O}$; the mean, $\overline{\ln \ec}^{\rm O}$, is plotted as a solid black curve. (On the residual scale, the mean is always zero.) The dashed, dotted-dashed, and dotted lines plot point-wise intervals for the full range, middle $95\%$, and middle $68.3\%$ of $10,000$ replicates sampled under the PCA-based Gaussian model. The light, dark, and darker gray areas give the same intervals for the $10,000$ transformed emissivity curves in $\emissox$. The correspondence between $\emissox$ and the PCA-based Gaussian model  is quite good, especially for the $68.3\%$ intervals.

\section{Statistical Model and Inference}
\label{sec:stats}

Here we describe how we combine the \Fe \ and \OHe \ spectra (tabulated in $\datafe$ and $\dataox$, respectively) to estimate the plasma parameters, $\log \Ne$, $\log \volume$, and $\logtemp$,
the background parameters $\paramfebg$ and $\paramoxbg$, and the parameters describing the \Fe \ and \OHe \ emissivities (using $\rvfe$ and $\rvox$ via the PCA-based model described in Section~\ref{sec:atomic_pca}).

\subsection{Accounting for uncertainty in the emissivities.}
\label{sec:emiss-uncerntainty} 

Following \cite{yu2018incorporating} we consider two strategies for accounting for emissivity uncertainty.  The pragmatic-Bayes approach, averages inference over the PCA-based prior distributions presented in Section~\ref{sec:atomic_pca}. Specifically, pragmatic-Bayesian inference is based on
\begin{equation}
    p(\theta \mid \datafe, \dataox, \rvfe, \rvox) 
    \ p(\rvfe, \rvox),
    \label{eq:pragbayes}
\end{equation}
where $\theta$ represents the source and background parameters associated with both the \Fe \ and \OHe \ spectra. The fully-Bayesian approach, on the other hand, is based on the full posterior distribution, 
\begin{equation}
    p(\theta \mid \datafe, \dataox, \rvfe, \rvox) 
    \ p(\rvfe, \rvox \mid \datafe, \dataox).
    \label{eq:fullybayes}
\end{equation}
Notice that the pragmatic-Bayesian approach does not use the spectral data to update the posterior distributions of the emissivities and leaves the uncertainty and estimation of the emissivities completely in the hands of the preliminary analysis conducted by atomic physicists. This approach avoids any biasing of the estimated emissivities that might stem from potential misspecification of the spectral models, systematic uncertainties associated with the spectral analysis, or peculiarities associated with the spectral data. By contrast, the fully-Bayesian analysis uses the spectral data to update and learn the emissivities. We generally expect both strategies to be more reliable than the standard practice of ignoring uncertainty in the emissivities. The fully-Bayesian approach generally provides smaller statistical errors than the pragmatic strategy in that it more fully utilizes the spectral data. The trade offs between the two methods are discussed and illustrated in \citet{lee2011accounting}, \citet{xu2014fully}, and \citet{yu2018incorporating}.   

Under either strategy, our Bayesian inference is based on a Monte Carlo (MC) sample from the target distribution, i.e., Equation~\ref{eq:pragbayes} for the pragmatic-Bayes approach and Equation~\ref{eq:fullybayes} for the fully-Bayesian approach. Obtaining an MC sample from the pragmatic-Bayesian target distribution is somewhat simpler because we can first sample $\rvfe$ and $\rvox$ directly from their respective prior distributions and then employ a Markov Chain Monte Carlo (MCMC) technique to sample the five-dimensional distribution $p(\theta \mid  \datafe, \dataox, \rvfe, \rvox)$. The fully-Bayesian strategy deploys MCMC to jointly sample $p(\theta, \rvfe, \rvox \mid \datafe, \dataox, \rvfe, \rvox)$. The dimension of this distribution depends on the number of principal components we deploy for the emissivity model.  Using $7$ and $6$ principal components for \Fe \ and \OHe, respectively, yields a $16$-dimensional fully-Bayesian posterior distribution. 

We employ two MCMC algorithms to obtain MC samples from the target posterior distributions,  a four-step Gibbs sampler with adaptive Metropolis (detailed in Appendix~\ref{sec:app-algs}) and a Hamiltonian Monte Carlo (HMC) sampler \citep{neal2011mcmc}. We implement HMC using the Stan computing platform\footnote{\url{https://mc-stan.org}}. Having two independent algorithms allows us to cross check and verify our numerical results. 

\subsection{Combining \Fe \ and \OHe \ analyses to estimate the plasma parameters} \label{sec:three_stage_analysis}

We conduct the spectral analysis in two stages, first analyzing $\datafe$ and then $\dataox$. To facilitate this, we separate the parameters into those that appear only in the model for $\datafe$, i.e., $\paramfe = (\rvfe, \paramfebg)$, those that appear only in the model for $\dataox$, i.e., $\paramox = (\rvox, \log \Ne, \paramoxbg)$, and those that appear in both models, $\paramoxfe = (\log \volume, \logtemp)$. (Here we mingle the plasma and the background parameters  together for simplicity.) 

Under the fully-Bayesian strategy, our primary aim is to study the posterior distribution, 
\begin{align}
p(&\paramox, \paramfe,  \paramoxfe \mid \dataox, \datafe) \nonumber \\
&\propto p(\dataox \mid \paramfe, \paramox, \paramoxfe, \datafe)   \cdot p(\paramfe, \paramox, \paramoxfe \mid \datafe)
\nonumber \\
&= p(\dataox \mid \paramox, \paramoxfe)  \cdot p(\paramox) \cdot p(\paramfe, \paramoxfe \mid \datafe), \label{eq:jointposterior_OxFe_v2}
\end{align}
where the proportional expression follows from Bayes theorem and the equality from the conditional independence of $\datafe$ and $\dataox$ given the parameters, that $\dataox$ does not depend on $\paramfe$, that $\datafe$ does not depend of $\paramox$, and from our assumption that $\paramfe$, $\paramox$, and $\paramoxfe$ are {\it a priori} independent, see Section~\ref{sec:priors}. 

Equation~\ref{eq:jointposterior_OxFe_v2} is the road map for our two-stage analysis:
\begin{align}
\text{Stage 1: } & \datafe \mid \paramfe,  \paramoxfe, \\
\text{Stage 2: }  & \dataox \mid \paramox, \paramoxfe.
\end{align}
In Stage~1 we obtain a MC sample from the last factor in Equation~\ref{eq:jointposterior_OxFe_v2}, the posterior distribution of $\paramfe$ and $\paramox$ given $\datafe$. In Stage~2, treating this posterior as a prior distribution, we recognize Equation~\ref{eq:jointposterior_OxFe_v2} as the product of the likelihood of $\dataox$ and two independent priors, one for $\paramox$ and one for $(\paramfe,\paramoxfe)$. The Stage~1 analysis uses $\datafe$ as the source of information for the plasma volume and temperature (i.e., for $\paramoxfe$). The resulting marginal posterior distribution for $\paramoxfe$ is subsequently used as a prior distribution, along with an independent prior for the plasma density, in the State~2 analysis of $\dataox$. As described in Section~\ref{sec:fe_model}, we construct a parametric approximation to $p(\log \volume, \logtemp \mid \datafe)$ in order to simplify its use as a prior distribution in Stage~2. 

We take the same two-stage approach when conducting a pragmatic-Bayesian analysis. An expression similar to that in Equation~\ref{eq:jointposterior_OxFe_v2} can be derived  by conditioning throughout on $\rvfe$ and $\rvox$ and multiplying by $p(\rvfe, \rvox)$. Both stages of the analysis are conducted following the pragmatic strategy. In Stage~1, for example, we sample $\rvfe$ from its prior distribution and then update the other parameters (i.e., $\log \volume$, $\logtemp$, and $\paramfebg$) via MCMC conditional on $\rvfe$.

\subsection{Prior Distributions} \label{sec:priors}

We assume that the model parameters are {\it a priori} independent, i.e., that the joint prior distribution can be written
\begin{align}
p(& \log \volume, \logtemp, \log \Ne, \rvfe, \rvox, \paramfebg, \paramoxbg) \nonumber\\
&= 
 p(\log \volume) \ p(\logtemp) \ p(\log \Ne) \ p(\rvfe) \ p(\rvox) \ p(\paramfebg) \ p(\paramoxbg).
\label{eq:prior}
\end{align}
Marginally, we assume $\log \volume$,  $\logtemp$, and $\log \Ne$
are uniformly distributed, $\rvfe$ and $\rvox$ follow  multivariate standard Gaussian distributions, and $\paramfebg$ and $\paramoxbg$ follow Gamma distributions,  specifically, 
\begin{align}
  p(\log \volume) &= \frac{1}{4} \quad \ \ \text{for } 30 \le \log \volume \le 34, \label{eq:prior_logalpha0} \\
  p(\logtemp) &= \frac{1}{2.1} \quad \ \ \text{for } 5.8 \le \logtemp \le 7.9, \label{eq:prior_alpha1} \\
  p(\log \Ne) &= \frac{1}{7} \quad \ \ \text{for } 8 \le \log \Ne \le 13,\\
  \paramfebg &\sim \operatorname{Gamma} (\text{shape}=0.5, \text{rate}=2), \\
\paramoxbg &\sim \operatorname{Gamma} (\text{shape}=0.5, \text{rate}=2), 
  \label{eq:prior_bgfe}
\end{align}
and the prior distributions for $\rvfe$ and $\rvox$ are given in Equations~\ref{eq:prior_rvfe} and \ref{eq:prior_rvox}.

\subsection{Stage 1: Analysis of \Fe \ Lines} \label{sec:fe_model}

The likelihood function for the Stage~1 \Fe \ spectral analysis is given by Equations~\ref{eq:likeli_so_pois} and \ref{eq:likeli_bg_pois}. Specifically, given $\paramfe$ and $\paramoxfe$, the source and background counts are Poisson for each $\wvl \in \mathcal{W}^{Fe}$,
\begin{align}
      \datafeso(\wvl) 
      & \distas{\rm{indep}} \operatorname{Poison} \big( \estfeso(\wvl; \paramfeso) + \estfecont(\wvl) + \paramfebg \big), \label{eq:likeli_fe_so_pois} \\
      \datafebg(\wvl) 
      & \distas{\rm{indep}} \operatorname{Poison} \big( \ratio \cdot \paramfebg \big). \label{eq:likeli_fe_bg_pois} 
\end{align} 
where $\paramfeso=(\log \volume, \logtemp,\rvfe)$. Combining the likelihood under Equations~\ref{eq:likeli_fe_so_pois} and \ref{eq:likeli_fe_bg_pois}, and the prior distributions, given in  Equations~\ref{eq:prior_logalpha0} -- \ref{eq:prior_bgfe}, via Bayes theorem, we obtain the joint posterior distribution, $p(\log \volume, \logtemp, \rvfe, \paramfebg \mid \datafe)$. 

Because we carry the joint posterior distribution of $\log \volume$ and $\logtemp$ forward to use as a prior distribution in Stage~2, we must obtain an analytical approximation. In the numerical results described in Sections~\ref{sec:fe17_posterior_summary}, we use two analytical models, specifically a multivariate Gaussian distribution and a multivariate $t$-distribution. We denote the analytical approximation $\hat{p}(\log \volume, \logtemp \mid \datafe)$. 

\subsection{Stage 2: Analysis of \OHe \ Lines} \label{sec:ox_model}

Again following Equations~\ref{eq:likeli_so_pois} and \ref{eq:likeli_bg_pois}, the likelihood function for the Stage~2 \OHe \ spectral analysis is determined by the Poisson models,
\begin{align}
      \dataoxso(\wvl) 
      & \distas{\rm{indep}} \operatorname{Poisson} \big( \estoxso(\wvl; \paramoxso) + \estoxcont(\wvl) + \paramoxbg \big), \label{eq:likeli_ox_so_pois} \\
      \dataoxbg(\wvl) 
      & \distas{\rm{indep}} \operatorname{Poisson} \big( \ratio \cdot \paramoxbg \big). \label{eq:likeli_ox_bg_pois} 
\end{align}
for each $\wvl \in \mathcal{W}^{O}$, where $\paramoxso= (\log \volume, \logtemp, \log \Ne,  \rvox)$; Equations~\ref{eq:likeli_ox_so_pois} and \ref{eq:likeli_ox_bg_pois} are both conditional on $\paramox$ and $\paramoxfe$.
We obtain the Stage~2 posterior distribution by combing this likelihood with the prior distributions given in Section~\ref{sec:priors}, but with the joint prior distribution for $\log \volume$ and $\logtemp$ replaced by $\hat{p}(\log \volume, \logtemp \mid \datafe)$ as obtained in the Stage~1 analysis. 

\section{Analysis} \label{sec:capella}
\subsection{Stage~1 -- Analysis of  \Fe\ Lines} \label{sec:fe17stage}
\subsubsection{\Fe \ Simulation Studies} \label{sec:fe17_sim_study}

We begin with a simulation study conducted to validate our models and methods. Our aim is to evaluate our method's ability to recover the ground-truth plasma parameters from \Fe \ counts. To this end, we generate several replicate sets of source and background counts using realistic values for the model parameters (i.e., the ground truth). 

The ground-truth parameter values are set equal to the posterior means obtained in our fully-Bayesian analysis of the Capella data, described in Section~\ref{sec:fe17_obs_study}, i.e., we set $\log \volume=31.17$, $\logtemp=6.75$, $\paramfebg=11.11$, and $\rvfe=(6.15,  5.38, -4.45, -7.20, 8.49,  3.54, 0.81)$ for the $J=7$ principal components.  The emissivity is computed via Equation~\ref{eq:pca_emiss_replicate} with $\rv$ set to this value of $\rvfe$. We simulate $30$ sets of source and background photon counts for each channel from the Poisson distributions in Equations~\ref{eq:likeli_fe_so_pois} and \ref{eq:likeli_fe_bg_pois}. Each replicate data set is fit with our proposed method using HMC via Stan to obtain a sample from the joint posterior distribution of all unknown model parameters. For each replicate, we obtain $1000$ draws by running the HMC sampler for $6000$ iterations and discarding the first $5000$ as burn-in.

\begin{figure}[!t]
\centering
\includegraphics[width=\linewidth]{ 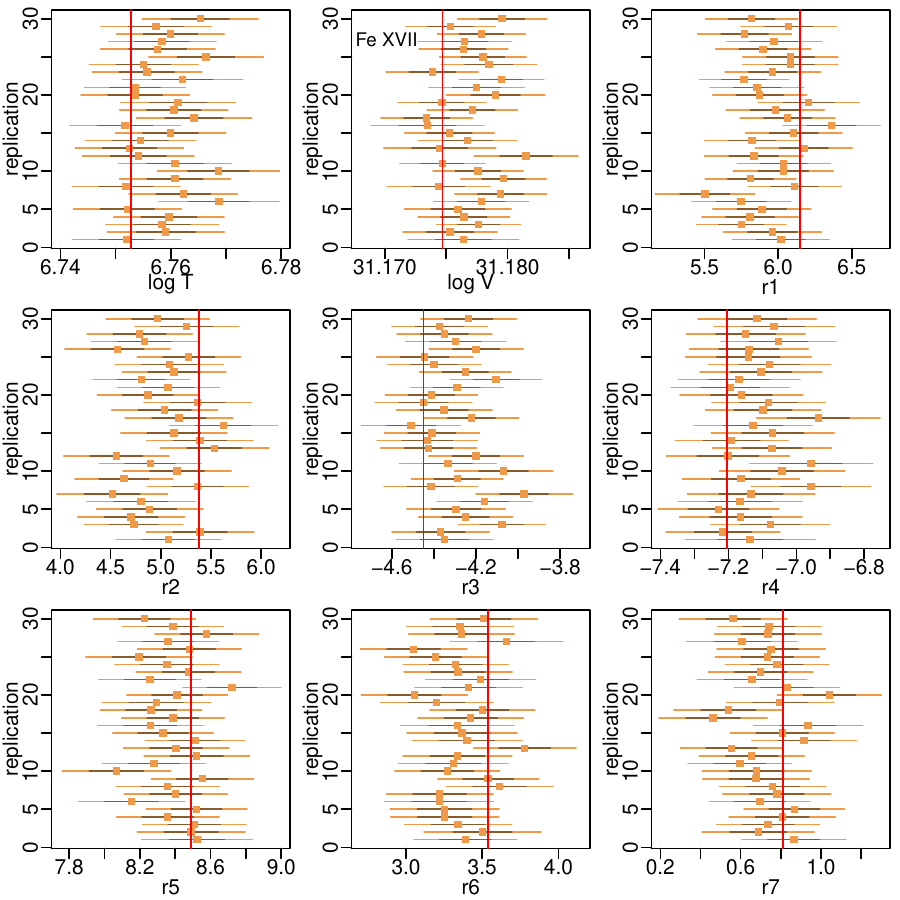}
\caption{Stage~1 \Fe\ Simulation Study. The posterior means (brown squares) along with $68\%$ and $95\%$ posterior intervals (dark and light brown horizontal bars) for each of the $30$ simulated \Fe\ datasets. The ground-truth parameter values are marked as red vertical lines. Apparent misalignment between some of the posterior means and the ground-truth values is an indication of a small-sample bias. \citet[][Figure~6.10 and 6.11]{yu2020multistage} numerically verified that this discrepancy dissipates as the exposure time increases.}
\label{fig:fe17_ctnm_sim_study_posterior_sle_CI}
\end{figure}

The results of the simulation study are presented in Figure~(\ref{fig:fe17_ctnm_sim_study_posterior_sle_CI}). Posterior means for each of the 30 simulated datasets are indicated by brown squares, with dark and light brown horizontal bars representing $68\%$ and $95\%$ posterior intervals, respectively. The ground-truth values of the model parameters are indicated by the red vertical lines. The coverage rates of the $68\%$ and the $95\%$ posterior intervals for all parameters are  approximately $68\%$ and $95\%$, respectively, an indication that our Bayesian methodology has desirable frequentist properties. Further, the profile of the expected photon counts (not shown) shows strong agreement with that of the simulated counts, and examination of the standardized residuals did not reveal any worrisome patterns or deviation from independent white noise. The fact that the posterior means (brown squares) of $\logtemp$, $\log \volume$, and some components of $\rvfe$ are on average misaligned with their ground-truth values is an indication of a small-sample bias. \citet[][Figure~6.10 and 6.11]{yu2020multistage} numerically verified that this apparent bias dissipates as the exposure time increases. Overall, the simulation results demonstrate the strength of our methodology for recovering the plasma parameters from observed data.

\subsubsection{Application to Observed Capella \Fe\ Counts} \label{sec:fe17_obs_study}
Turning to the analysis of Capella, as discussed in Section~\ref{sec:preprocessing}, for an observed dataset from grating MEG and positive order, denoted by {\tt amp}, we consider $\iH = 100$ channels for $\iL=13$ spectral lines. The weak Fe line at 16.004~\AA, which is also blended with an \ion{O}{8} line, has been excluded from the analysis (see Section~\ref{sec:preprocessing}).

We fit the model summarized in Section~\ref{sec:fe_model} using the four-step Gibbs sampler\footnote{We obtain an MC sample of size 2 million from a single chain obtained with the four-step Gibbs Sampler (for both the pragmatic and fully-Bayesian approaches).  As described in Appendix~\ref{sec:app-algs}, the first $1000$ iterations are used to tune an adaptive Metropolis update. The first half of the remaining iterations is discarded as burn-in.} 
described in Appendix~\ref{sec:app-algs} for the pragmatic-Bayesian analysis and using HMC with Stan\footnote{We ran four independent HMC chains with randomly-selected starting values, each for 6000 iterations. The first 5000 iterations of each chain are discarded as burn-in.  Multiple chains with dispersed starting values are run to check that consistent convergence is achieved.} for the fully-Bayesian analysis. (We compare the performance of the four-step Gibbs sampler with that of HMC in the fully-Bayesian analysis in Appendix~\ref{app:compare_algs}.)

Figure~\ref{fig:fe17_ctnm_pragBfullB} compares the marginal posterior densities of the model parameters obtained from the pragmatic (amber) and the fully (green) Bayesian analyses.  As expected, incorporating the prior uncertainties of the emissivities under the pragmatic-Bayesian analysis yields wider marginal posterior densities as compared with the fully-Bayesian analysis. The latter uses the data to narrow the posterior uncertainty in $\rvfe$ which in turn is expected to narrow the uncertainty in the plasma parameters. For $\logtemp$ and $\log \volume$, there is little overlap between the two posterior densities. 

This indicates that these parameters are particularly sensitive to uncertainty in the emissivities and that the observed line counts add considerable information regarding likely values of the emissivities. 

The Gaussian approximation that we use for the prior distribution on the emissivities (see Section~\ref{sec:atomic_pca}) includes appreciable probability beyond the range of the ensemble, $\emissfe$. This allows the fully-Bayesian posterior distributions of some components of $\rvfe$ to be centered as far as eight standard deviations from their prior means, e.g., $r_4$ and $r_5$ in Figure~\ref{fig:fe17_ctnm_pragBfullB}. The effect of this on the 13 \Fe\ line emissivities is illustrated in Figure~\ref{fig:fe17_emis_default_fitted}, where the data-driven (fully-Bayesian) posterior means (green lines) differ substantially from the default CHIANTI values (red lines) in some cases, particularly for the lines at $13.89$, $12.12$, and $15.01\mathring{\text{A}}$ where the posterior means are sometimes outside of the range of $\emissfe$. The data-driven estimates of the atomic emissivities for the $15.01$ and $12.12 \mathring{\text{A}}$ lines are about $30\%$ smaller and about $100\%$ larger, respectively, than their default CHIANTI values. The estimated emissivities for the other lines are much closer to their default values.

\begin{figure}[!t]
\centering
\includegraphics[width=\linewidth]{ 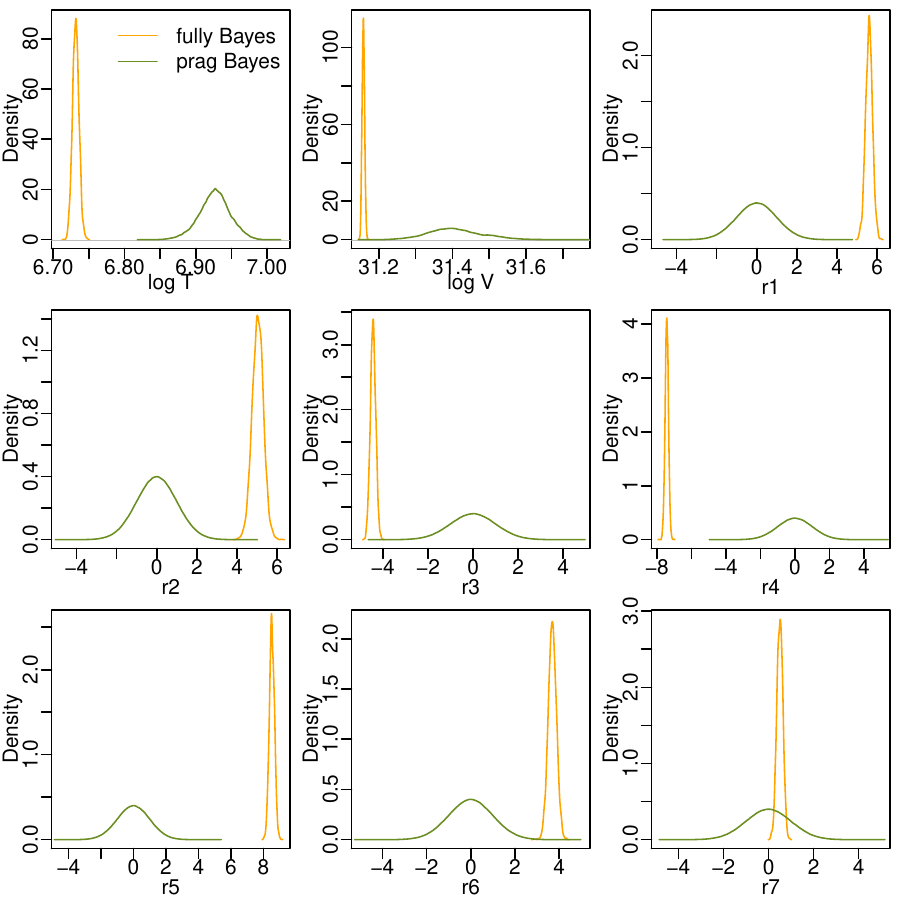}
\caption{Stage~1 \Fe\ Analysis:
Comparison of the posterior distributions of the model parameters obtained with the pragmatic (green) and the fully (amber) Bayesian analyses.}
\label{fig:fe17_ctnm_pragBfullB}
\end{figure}

\begin{figure*}[t]
\centering
\includegraphics[width=0.85\linewidth]{ 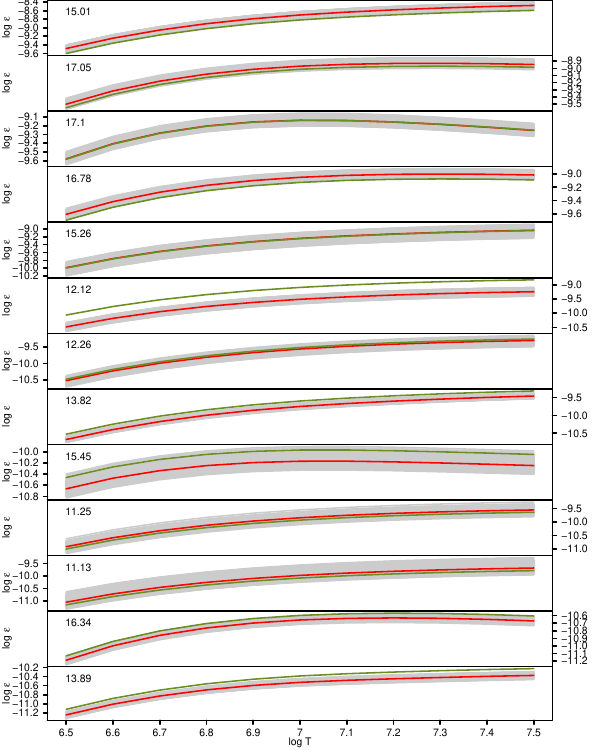}
\caption{
Emissivities of the 13 \Fe \ lines considered ($\log$ scale). The grey lines represent the $1000$ realizations of the CHIANTI atomic data in $\emissfe$. The red curve is the default value from the CHIANTI and the green curve is the emissivities preferred by the data (i.e., the posterior mean) obtained with the fully-Bayesian analysis.}
\label{fig:fe17_emis_default_fitted}
\end{figure*}

\begin{figure}[!t]
\centering
\includegraphics[width=\linewidth]{ 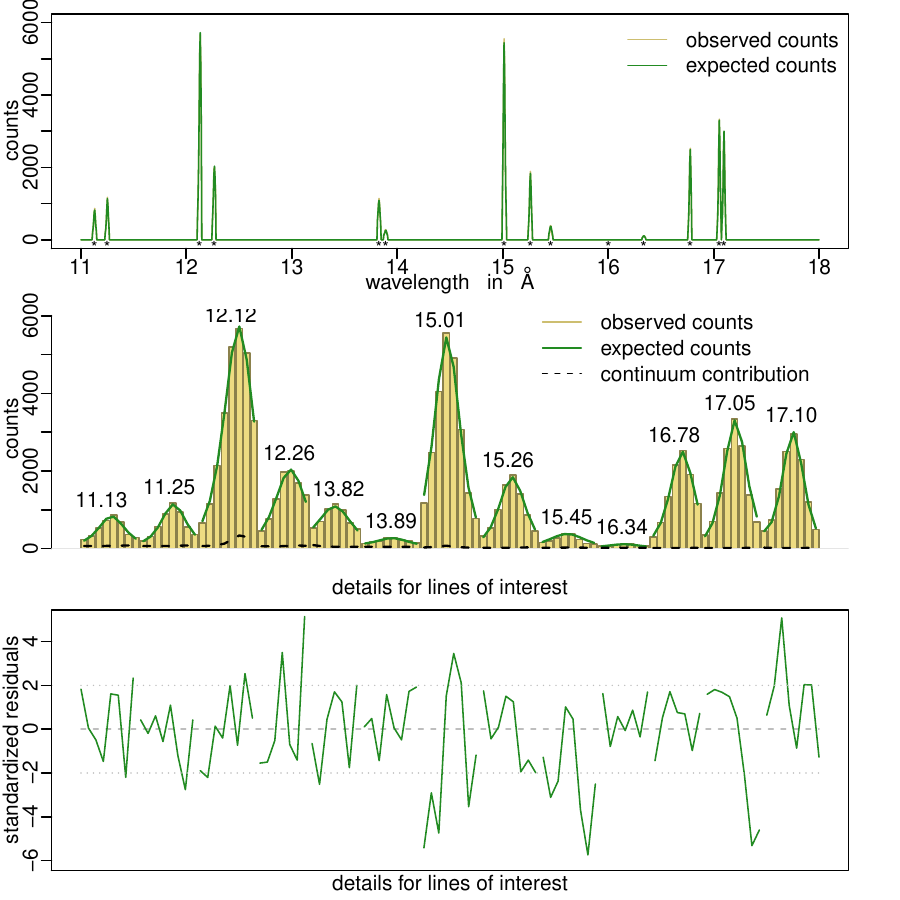}
\caption{Comparison of the expected \Fe\ line profiles with the observed counts from {\tt amp}. The top panel compares the expected photon counts computed with the model parameters set to their fully-Bayesian posterior means (green) with the observed counts (yellow). Lines of interest are marked '*'. The middle panel magnifies the top panel by focusing on the wavelength ranges covered by the lines of interest. It is constructed by removing wavelength gaps not included in the analyses to more closely compare the spectral profiles. The wavelength are reported at the top of each line. The dashed black curve represents the contribution of the continuum counts to the spectrum. The bottom plot represents the standardized residuals, i.e., the difference between the observed counts and their expected value under the model divided by the square root of the expected counts (since the counts are modeled as Poisson).}
\label{fig:fe17_spectrum_ctnm}
\end{figure}

The top panel of Figure~\ref{fig:fe17_spectrum_ctnm} compares the observed counts (yellow) in the source exposure with expected counts (green) obtained using the fully-Bayesian posterior means of the model parameters. The good alignment between the two is magnified in the middle panel which omits gaps in wavelength (i.e., those wavelengths not considered in our analyses) to allow examination of details of the spectral profiles. The dashed black curve represents the contribution of the continuum counts to the spectrum, which is about $51$ counts per channel on average with a larger contribution at shorter wavelengths.
Standardized residuals ($\frac{\textrm{Data-Model}}{\sqrt{\textrm{Data}}}$) are shown in the lower panel, with the majority of the points falling within $\pm$2.  Notice, however, that the residuals for individual lines tend to have a characteristic triangular shape (most obviously at 15.01~\AA), which suggests that the adopted LRF (Equation~\ref{eq:lrf}) is too broad compared to the true LRF, and that a better match to the observed profiles could be obtained with $\nu<4$.  However, the residuals also indicate that the total fluxes in the lines are estimated to a precision of $approx$5\%, so we deem it unnecessary to try to improve the fits further.

\subsubsection{Results Based on Capella \Fe\ Line Counts} \label{sec:fe17_posterior_summary}

Our estimates of the plasma parameters based on their 
fully-Bayesian posterior samples are $\log \volume = 31.158 \pm 0.003$ and $\logtemp = 6.732 \pm 0.005$. As discussed in Section~\ref{sec:fe_model}, we carry the marginal posterior distribution, $p(\log \volume, \logtemp \mid \datafe)$, from Stage~1 forward as a prior distribution for the plasma parameters in Stage~2. To facilitate this, we consider two analytic approximations: a bivariate Gaussian distribution and a bivariate $t$-distribution with 4 degrees of freedom. The functional form of the univariate $t$-distribution is given in Equation~\ref{eq:lrf}.

\begin{figure}[!t]
\centering
\includegraphics[width=\linewidth]{ 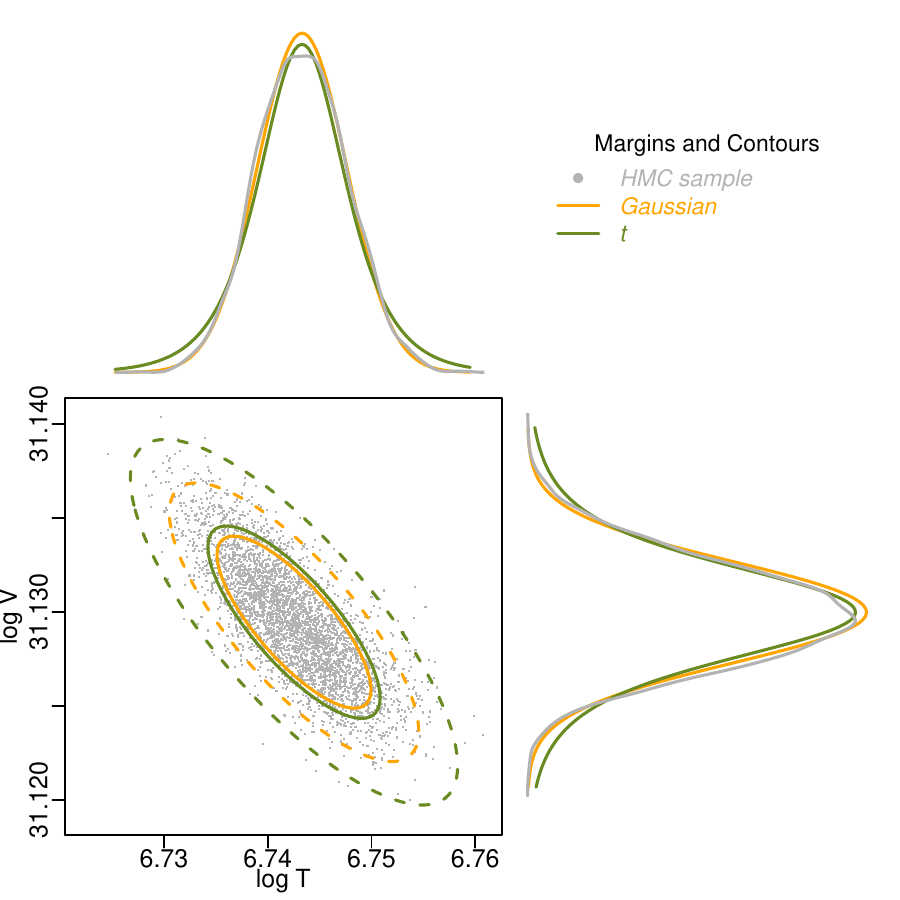}
\caption{Comparison of the Gaussian and $t$ approximations to the Stage~1 \Fe \ fully-Bayesian posterior distribution of $\log \volume$ and $\logtemp$. Lower left: $68\%$ (solid) and $95\%$ (dashed) posterior contours are plotted for the Gaussian (amber) and $t$ (green) approximations and compared with an MC posterior sample plotted in grey. 
The corresponding marginal distributions are plotted along the top and the right (using the corresponding axes).
}
\label{fig:fe17_posterior_summary_ctnm_contour}
\end{figure}

The approximations are compared in Figure~\ref{fig:fe17_posterior_summary_ctnm_contour} via their $68\%$ (solid) and $95\%$ (dashed) joint probability contours and via their marginal distributions, with amber representing the Gaussian approximation and green the $t$-approximation. An MC sample of $(\log \volume, \logtemp)$ from the fully-Bayesian posterior is plotted in grey. The contours and marginal distributions of the $t$-approximation are wider than those of the Gaussian approximation, owing to the heavier tails of the $t$ distribution. (Note particularly the 95\% contours of the joint distribution.) This is by design as the $t$ approximation is meant to be conservative. When carrying the approximation forward to Stage~2, a conservative approximation allows for possible systematic errors/differences between the \Fe\ and \OHe\ datasets.  Thus, we use the bivariate $t$-approximation in our Stage~2 analysis of \OHe. \cite{yu2020multistage} has also checked the sensitivity of the results to the specification of the prior distribution by comparing a two dimensional $t$-distribution with a two dimensional Gaussian distribution and two independent uniform distributions on $(\log \volume, \logtemp)$.

\subsection{Stage~2 -- Analysis of \OHe\ Lines}\label{sec:oxstage}
\subsubsection{\OHe \ Simulation Studies}
We conduct a second simulation study to validate our Stage~2 (\OHe) analysis. As in the simulation described in Section~\ref{sec:fe17_sim_study}, we generate several replicate sets of source and background counts using realistic values for the model parameters (i.e., the ground truth). Further, the values of the temperature and the volume are assumed to be the same in both stages. 

The ground-truth parameter values are set to $\log \volume=31.17$, $\logtemp=6.75$, $\log \Ne=9.43$, $\paramoxbg=0.93$, and  $\rvox=(-1.18,  0.61, 2.36, 0.77,  2.92, -0.77)$ for the $J=6$ principal components. These values are obtained from an initial fit of the observed \OHe\ line counts, except that the ground-truth of $\log \volume$ and $\logtemp$ are the same as in the \Fe \ simulation study. We simulate $30$ sets of source and background \OHe \ photon counts for each channel using the Poisson distributions in Equations~\ref{eq:likeli_ox_so_pois} and \ref{eq:likeli_ox_bg_pois}. The $t$ approximation to $p(\log \volume, \logtemp \mid \datafe)$ described in Section~\ref{sec:fe17_posterior_summary} is used as the prior distribution for $(\log \volume, \logtemp)$ 

\begin{figure}[!t]
\centering
\includegraphics[width=\linewidth]{ 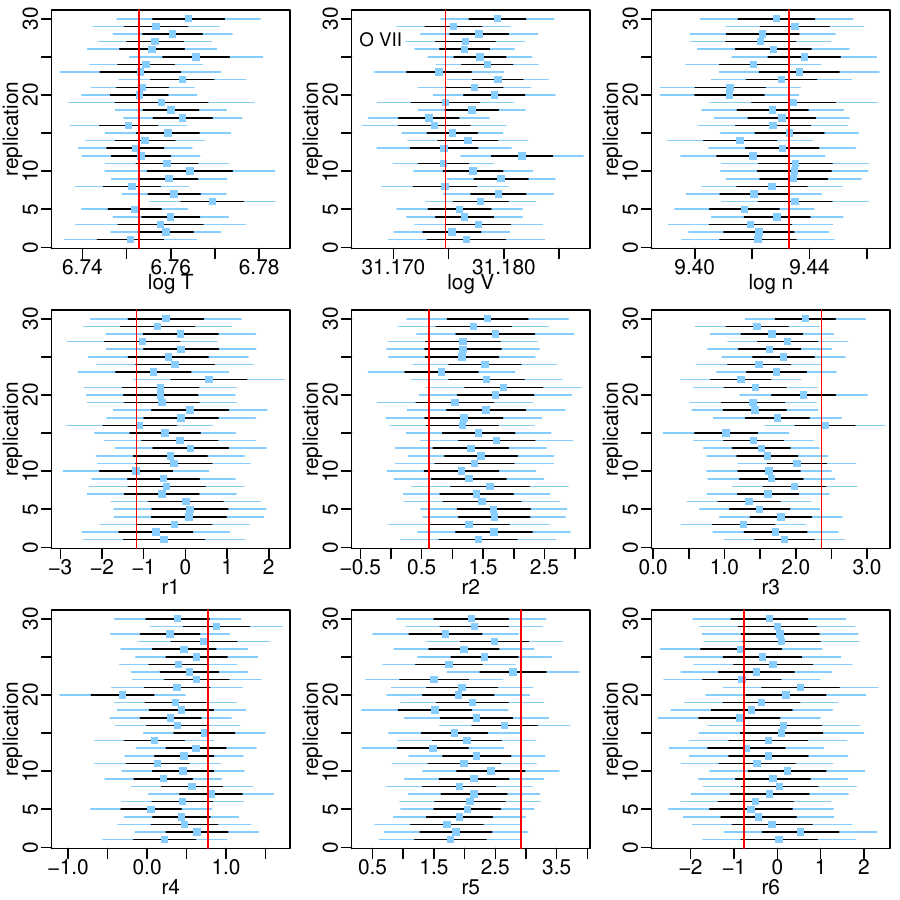}
\caption{Stage~2 \OHe\ simulation study. The posterior means (blue squares) along with $68\%$ and $95\%$ posterior intervals (black and light blue horizontal bars) for each of the $30$ simulated \OHe\ datasets. The ground-truth parameter values are marked as red vertical lines. The apparent small-sample bias diminishes as exposure time increases \cite[][Figure~6.10 and 6.11]{yu2020multistage}}
\label{fig:ox_sim_t_CIs_1time}
\end{figure}

We again use HMC via Stan for model fitting to obtain a sample from the fully-Bayesian posterior distribution described in Section~\ref{sec:ox_model} for each of the 30 simulated sets of \OHe \ line counts.\footnote{For each replicate data set, we ran four independent HMC chains with random starting values, each for $4000$ iterations. The first $3000$ iterations of each chain are discarded as burn-in.}
The results of the simulation are presented in Figure~\ref{fig:ox_sim_t_CIs_1time}.  Posterior means for each of the 30 simulated datasets are indicated by blue squares, with dark and light blue horizontal bars representing the $68\%$ and $95\%$ posterior intervals, respectively. The ground-truth values of the model parameters are indicated by red vertical lines. As in the \Fe\ simulation study, the apparent bias attenuates with large exposure times \citep[][Figures~6.10 and 6.11]{yu2020multistage}.

\subsubsection{Application to Observed Capella \OHe\ Counts using \Fe-based Prior Distribution}

As discussed in Section~\ref{sec:preprocessing}, our observed \OHe\ data are comprised of source and background exposures in $\iH=112$ channels corresponding to $\iL=6$  spectral lines. We fit the model summarized in Section~\ref{sec:ox_model} using the four-step Gibbs sampler\footnote{The four-step Gibbs sampler was implemented as in the \Fe\ pragmatic-Bayesian analysis, but run for only 1 million iterations.} for the pragmatic-Bayesian analysis and using HMC with Stan\footnote{HMC was implemented as in the \Fe\ fully-Bayesian analysis, but each chain was run for only 4000, half of which were discarded as burn-in.} for the fully-Bayesian analysis. 

\begin{figure}[!t]
\centering
\includegraphics[width=\linewidth]{ 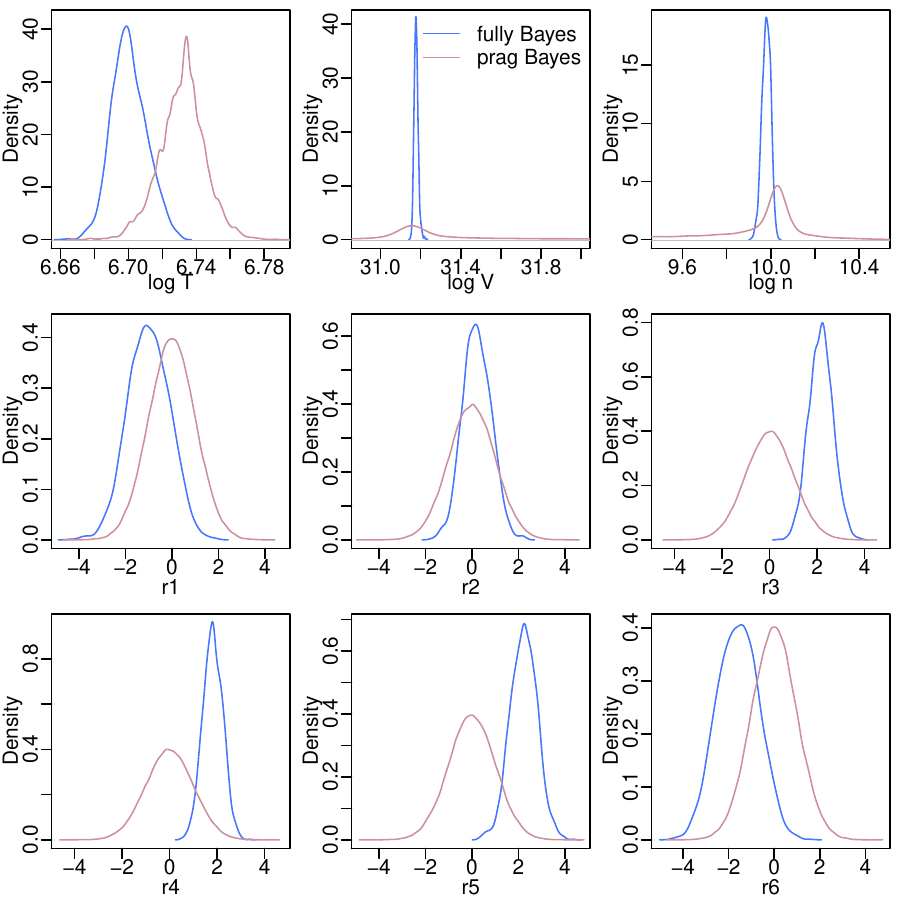}
\caption{Stage~2 \OHe\ analysis: comparison of the posterior distributions of the model parameters obtained with the pragmatic (amber) and the fully (blue) Bayesian analysis. }
\label{fig:ox_pragBfullB}
\end{figure}

\begin{figure*}
\centering
\includegraphics[width=0.85\linewidth]{ 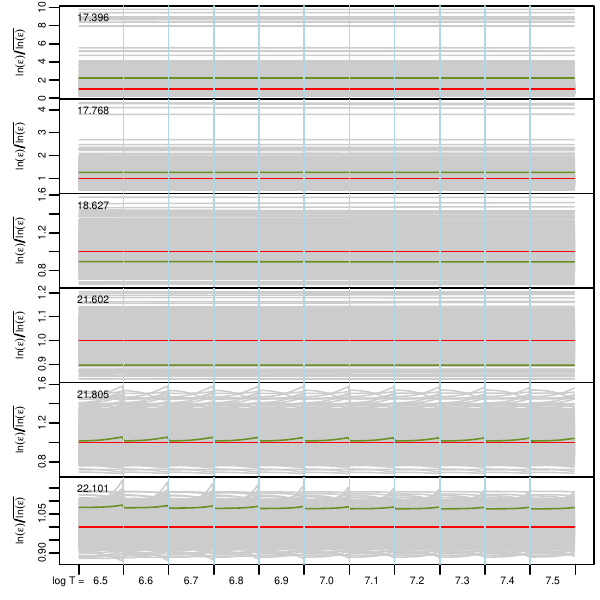}
\caption{Emissivitiy ratios (relative to the default) for the six \OHe \ lines. The grey lines represent the ratios of the $10,000$ realizations of the \OHe \ CHIANTI atomic data. The red curve corresponds to the default CHIANTI value (whose ratio with itself is one). The green curve is the data-driven estimate, i.e., the posterior mean of the emissivity curve under the fully-Bayesian analysis. The emmissivities are plotted as a function of $\log \Ne$ and $\logtemp$. In each panel, the horizontally-arranged blocks correspond to the values of the equally-spaced temperature grid, $\logtemp \in (6.5,7.5)$. Within each $\logtemp$ block, $\log \Ne$ increases from left to right in the interval $(9,11)$.}
\label{fig:ox_emis_default_fitted}
\end{figure*}

Figure~\ref{fig:ox_pragBfullB} compares the marginal posterior distributions obtained with the pragmatic and fully-Bayesian techniques for $\logtemp$, $\log \volume$, $\paramfebg$, $\log \Ne$, and each component of $\rvox$. The fully-Bayesian posterior distributions are narrower for $\log \Ne$ and $\log \volume$ and those of $\logtemp$, $\log \volume$, and $\log \Ne$, are shifted relative to their pragmatic counterparts, indicating that these parameters are sensitive to the emissivities (that are learned from the data with the fully-Bayesian approach). The posterior and prior distributions of the components of $\rvox$ are more aligned than those of $\rvfe$ (compare Figures~\ref{fig:fe17_ctnm_pragBfullB} and \ref{fig:ox_pragBfullB}), although the prior and posterior means differ by about three standard deviations in some cases, e.g., $\rvox_3$ and $\rvox_5$. Thus, there are cases where the data-driven fully-Bayesian estimates of the emissivities deviates appreciably from the CHIANTI default, e.g., line $21.602 \ \mathring{\text{A}}$, see Figure~\ref{fig:ox_emis_default_fitted}. 

\begin{figure}[!t]
\centering
\includegraphics[width=\linewidth]{ 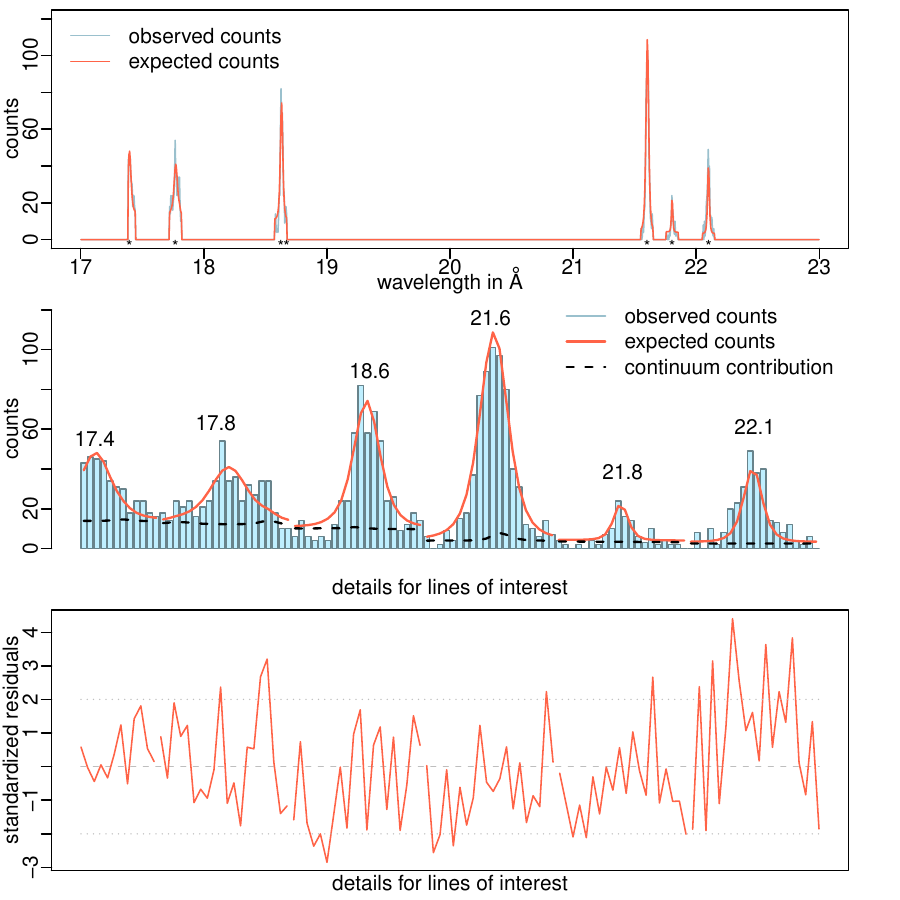}
\caption{
Comparison of the expected \OHe\ line profiles with the observed  counts from {\tt amp}. The three panels are the same as in Figure~\ref{fig:fe17_spectrum_ctnm}, except that expected total counts are indicated with a red curve and  observed counts in blue. 
}
\label{fig:ox_spectrum_t}
\end{figure}

Our estimates of the plasma parameters based on the two-stage fully-Bayesian posterior distributions are $\log \volume = 31.178 \pm 0.010$, $\logtemp = 6.701 \pm 0.010$, and $\log \Ne = 9.980 \pm 0.020$.  Figure~\ref{fig:ox_spectrum_t} replicates the model diagnostics given for the Stage~1 \Fe\ analysis in Figure~\ref{fig:fe17_spectrum_ctnm} for the Stage~2 \OHe\ analysis. We again find a good match between the observed counts and the expected counts under the model (computed using the posterior means of the model parameters). Here, the contribution of the continuum counts in the spectrum is about $6$ counts per channel on average.  The bottom panel of Figure~\ref{fig:ox_spectrum_t} shows that about $82\%$ of the standardised residuals are within the $95\%$ confidence interval, $(-2, 2)$, validating our model fit. 

\section{Discussion}\label{sec:discuss}

\subsection{Interpreting the estimated $\logtemp$ and $\Ne$}

{Stellar coronal EMs are known to be multi-thermal, and Capella is no exception.  However, repeated analyses and observations have shown that there is a strong excess in EM around $\logtemp\approx6.8$ (Brickhouse et al.\ 2000, Argiroffi et al.\ 2006), which justifies our approximating the DEM by isothermal plasma.  Nevertheless, we caution against overinterpreting the estimates found here, especially for $\logtemp$.  Our goal is to demonstrate the magnitude of the effect of atomic data uncertainties, whose main effect is to broaden the uncertainty in the estimates.  Note that in the absence of accounting for such systematic uncertainties, the uncertainties are driven solely by statistical fluctuations, which contribute $<0.3$\% to the error, or $<0.02$~dex and are obviously unrealistic.  As we see from Figure~\ref{fig:ox_pragBfullB}, the uncertainty in $\logtemp$ is $\approx$0.04~dex and that in $\Ne$ is $\approx$0.1~dex. }

\subsection{Advantages and Limitations}

We have developed a method that integrates X-ray spectral analysis and atomic line emissivity uncertainties to compute plasma parameters like temperature and number density.  While the structure of the algorithm is complex, due to the inclusion of diverse streams of data flow, and the computational cost is high, this method codifies the foundation of a process that allows for a flexible approach to high-resolution gratings analysis.  Our method is designed to be extensible to new datasets, and we demonstrate here that it produces estimates that are reasonable.

In order to focus the analysis on the specifics of the algorithm, we use a simplified model of the emission, i.e. adopted an isothermal approximation. However, several previous studies have found that the emission from Capella is well described by a near-isothermal distribution, and indeed our analysis results are consistent with previous estimates.

The plasma number density estimate relies on the density- and temperature-sensitive lines of \OHe\, for which we strongly constrain the temperature using a prior-stage analysis of the density-insensitive lines from \Fe.  Each stage of the analysis is thus insensitive to uncertainties in ionization balance.

The atomic emissivity uncertainties are computed here through physically motivated methods.  While the sets of emissivity tables that describe the deviations can be resource heavy, reducing their dimensionality with PCA allows us to interpolate within this high-dimensional sparsely sampled space.

The observational dataset we use is of Capella, with \chandra\ data combined across several years of observations to have high S/N.  Capella is a steady, but not invariable, source. This has the direct consequence that statistical uncertainties are relatively small, and we are able to explore the impact of systematic effects arising from atomic emissivity uncertainties. We do not account for either intrinsic source variability or calibration uncertainties here, but we compute the plasma parameters for different grating and detector combinations, with the resulting scatter consistent with calibration and variability.

While we use \chandra\ gratings spectral data here, our method easily generalizes to future high-resolution calorimeter spectra such as those expected from 
XRISM \citep{2022arXiv220205399X,2022SPIE12181E..1SI}, Athena \citep{2022arXiv220814562B}, Arcus \citep{2020SPIE11444E..2CS}, LEM \citealt{2022arXiv221109827K}, etc.

\subsection{Comparison across grating datasets}\label{sec:other_gratings}

\begin{figure}
    \centering
    \includegraphics[width=0.9\linewidth]{ 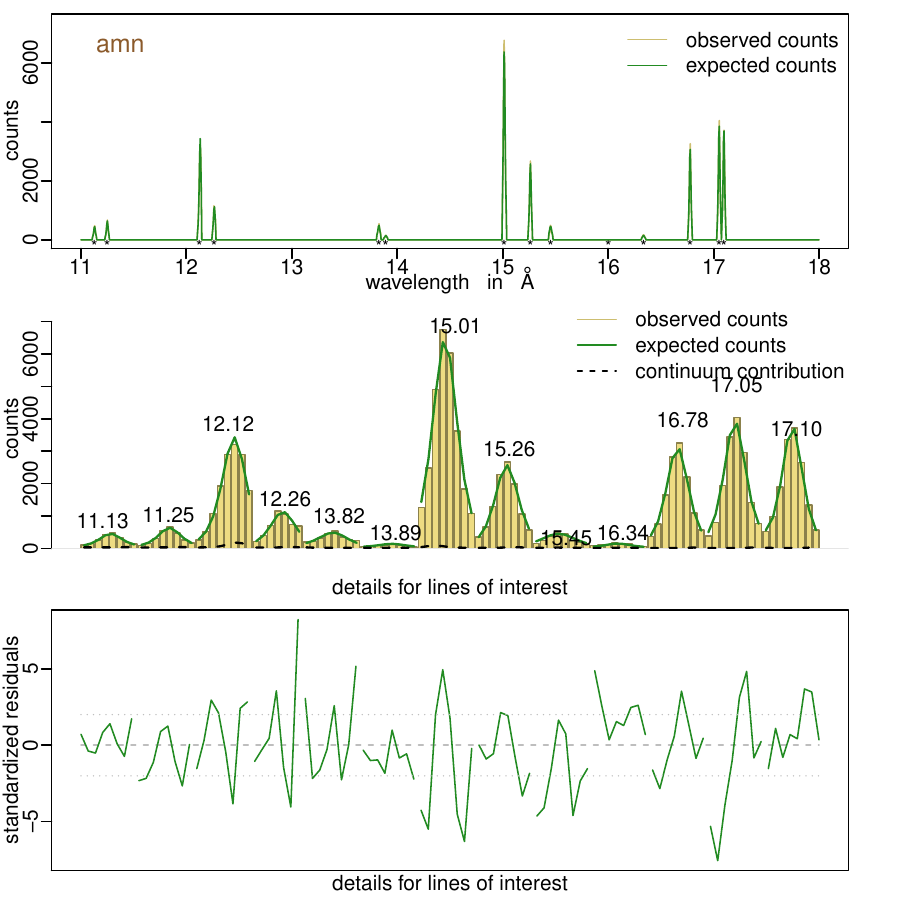}
    \includegraphics[width=0.9\linewidth]{ 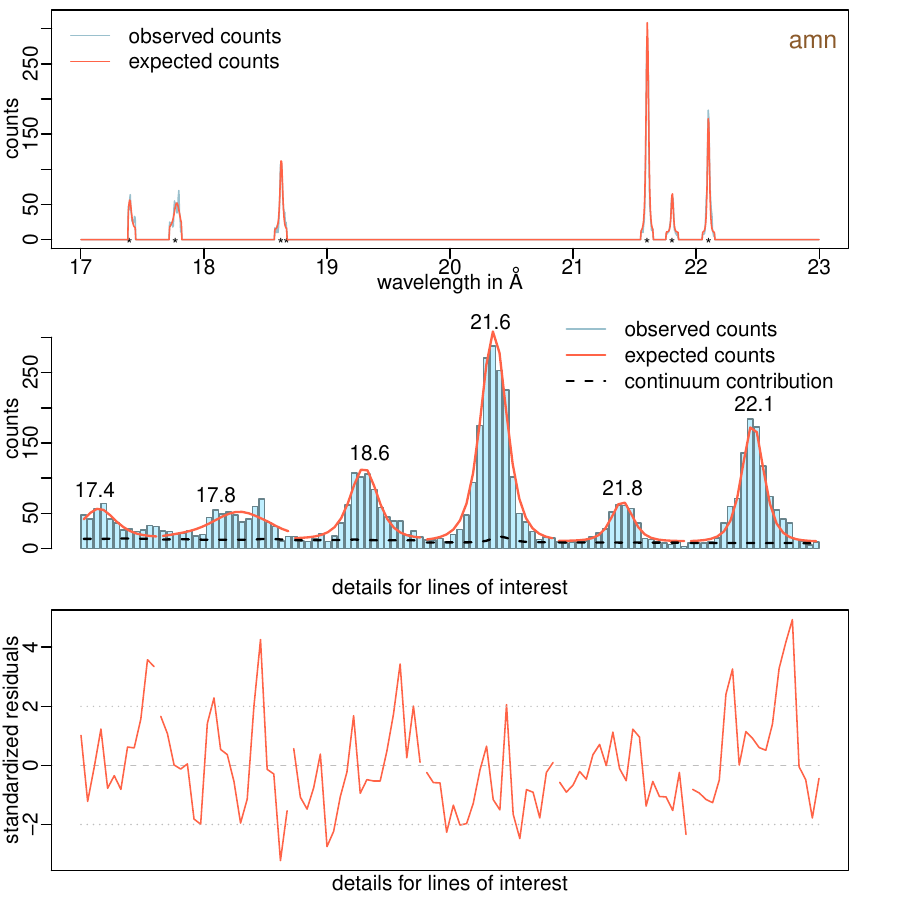}
    \caption{Summary of the analysis of the {\tt amn} dataset.  Upper three panels depict the analysis of \Fe\ as in Figure~\ref{fig:fe17_spectrum_ctnm}, and the bottom three panels depict the analysis of \OHe\ as in Figure~\ref{fig:ox_spectrum_t}.}
    \label{fig:amncombo}
\end{figure}

\begin{figure}
    \centering
    \includegraphics[width=0.9\linewidth]{ 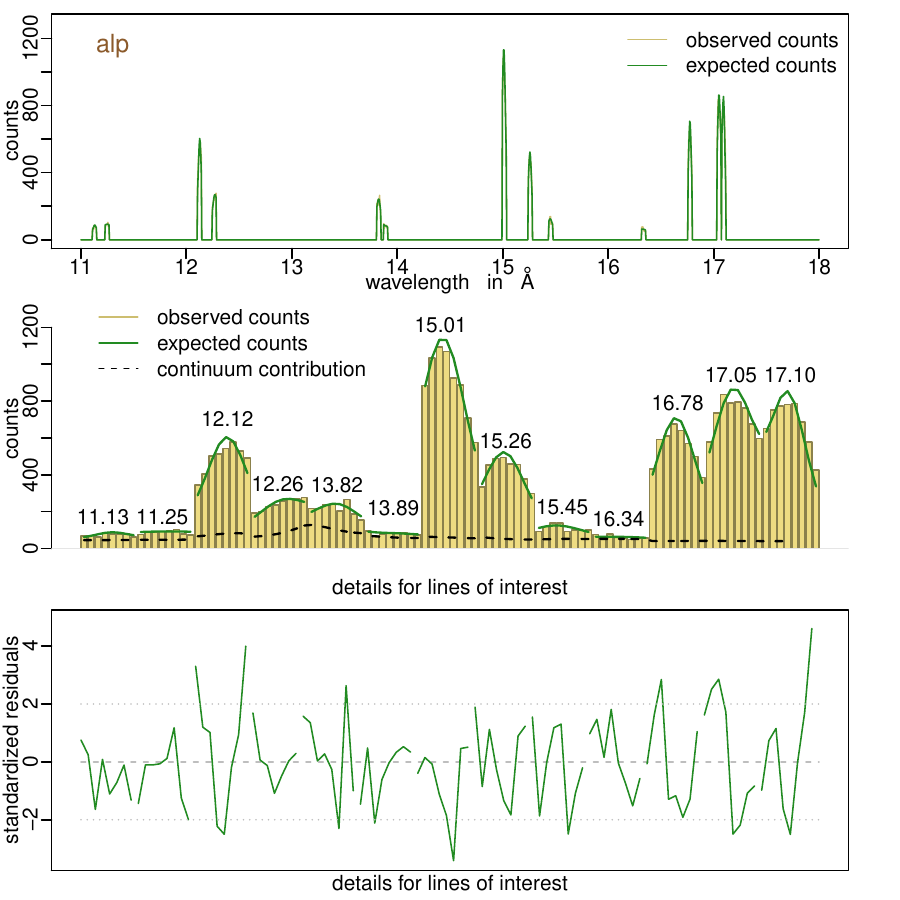}
    \includegraphics[width=0.9\linewidth]{ 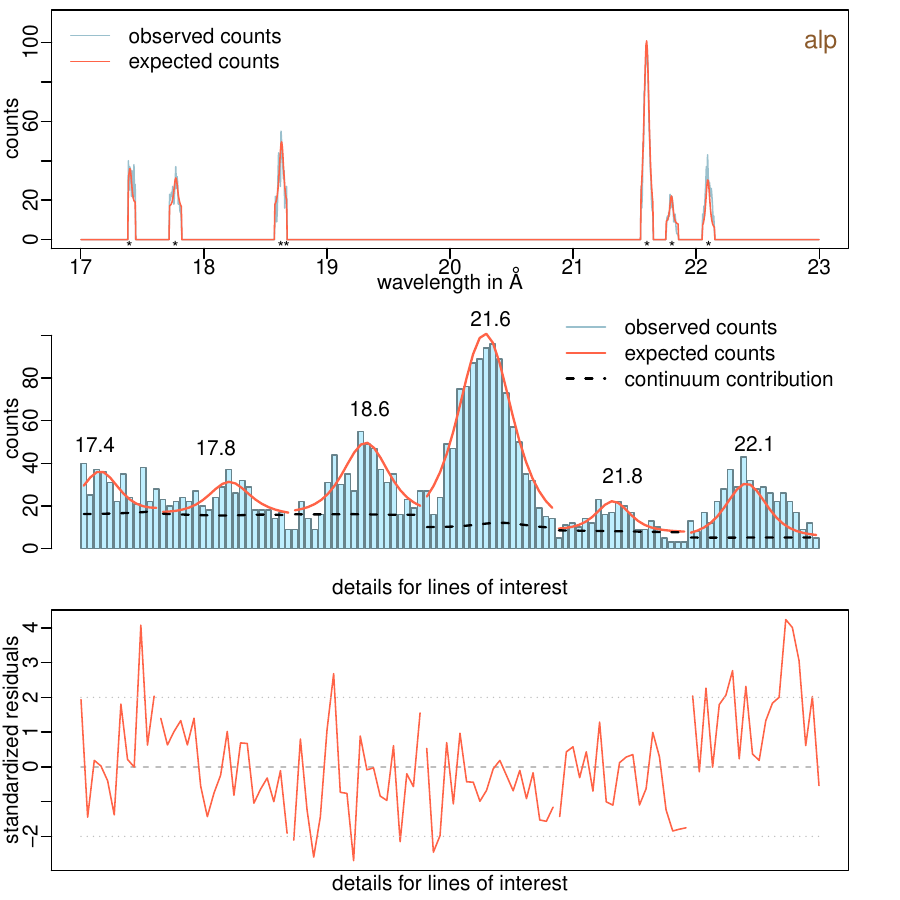}
    \caption{Summary of the analysis of the {\tt alp} dataset, with a similar arrangement as in Figure~\ref{fig:amncombo}.}
    \label{fig:alpcombo}
\end{figure}

\begin{figure}
    \centering
    \includegraphics[width=0.9\linewidth]{ 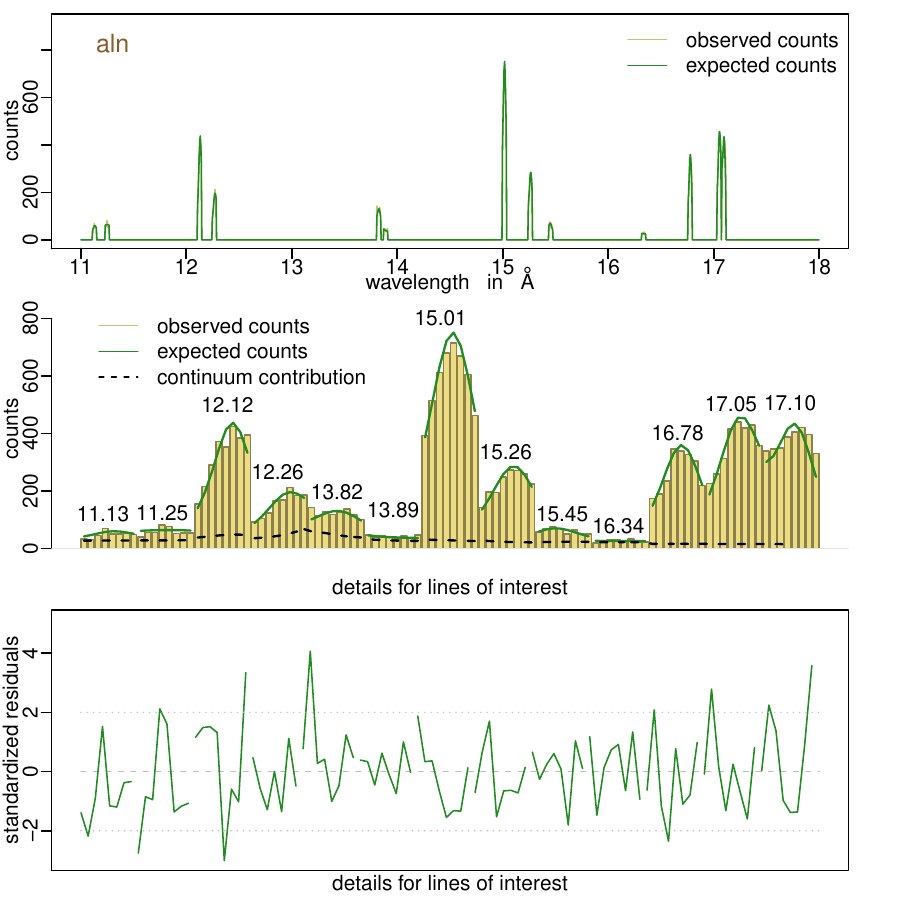}
    \includegraphics[width=0.9\linewidth]{ 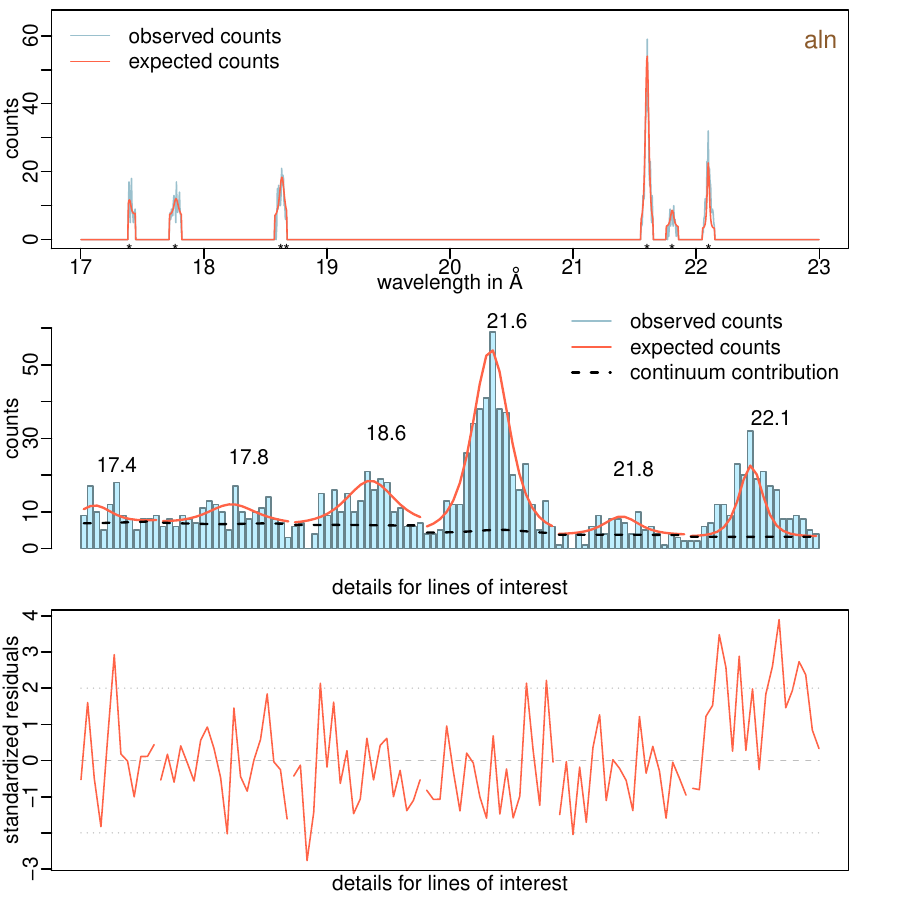}
    \caption{Summary of the analysis of the {\tt aln} dataset, with a similar arrangement as in Figure~\ref{fig:amncombo}.}
    \label{fig:alncombo}
\end{figure}

The \chandra\ observations of Capella form a unique high spectral resolution dataset with very long exposures (see Table~\ref{tab:chandra_capella}). Because of the huge number of photons that are observed, statistical noise has a negligible contribution to the uncertainty intervals compared to sources of systematic error such as atomic emissivity uncertainties, calibration uncertainties, and approximations in the astrophysical model of the coronal structure.  Above, we have focused primarily on characterizing the effects of atomic emissivity uncertainties using the {\tt amp} dataset.  There are other sets that can be used, such as the {\tt amn} (obtained simultaneously with {\tt amp}), as well as {\tt alp} and {\tt aln}, each of which is obtained simultaneously with the other. These datasets provide a natural opportunity to explore the stability of our algorithm and assess how many systematic effects still remain. We show the results of the analyses carried out the same way as for {\tt amp} on these other datasets in Figures~\ref{fig:amncombo}, \ref{fig:alpcombo}, and \ref{fig:alncombo}. These figures demonstrate that the individual line profiles are well fit, and the residuals are well behaved for all cases. 

\begin{figure*}[!htbp]
\includegraphics[width=0.32\linewidth]{ 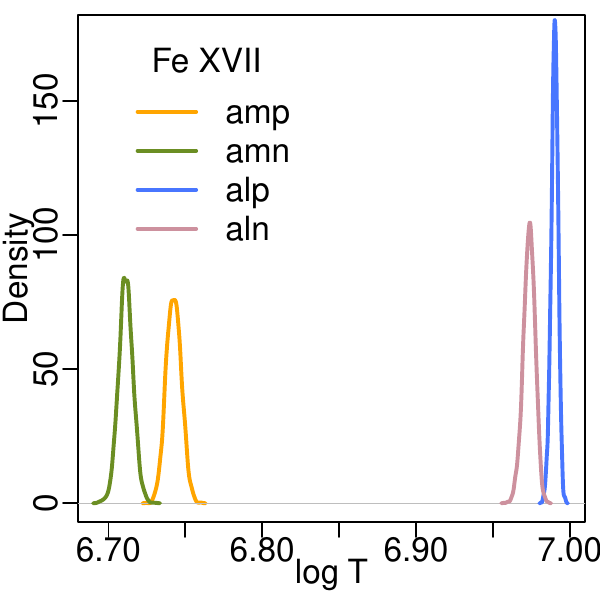}
\includegraphics[width=0.32\linewidth]{ 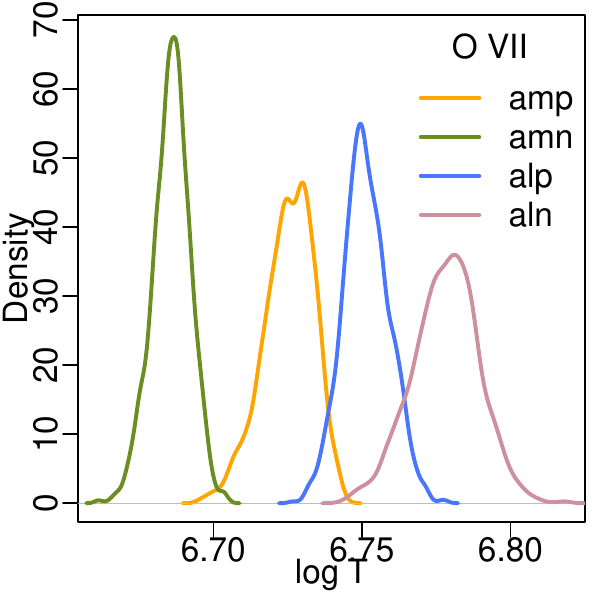}
\includegraphics[width=0.32\linewidth]{ 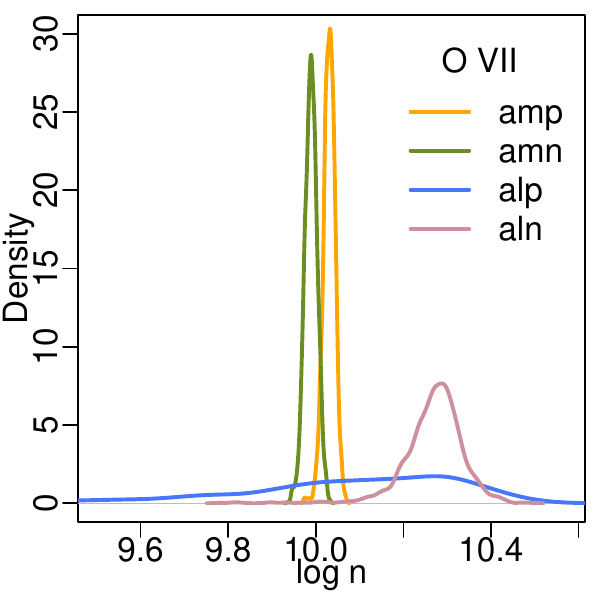}
\caption{Comparison of estimated plasma temperatures and densities over different datasets.  The {\tt amp}, {\tt amn}, {\tt alp}, and {\tt aln} analyses are shown as orange, green, blue, and pink curves respectively.  The left panel shows the posterior density estimates of the plasma temperature $\logtemp$ resulting from the first stage analysis of \Fe\ data.  The middle and right panels show the plasma density $\log \Ne$ and plasma temperature $\logtemp$ after the second stage analysis of \OHe\ data.}
\label{fig:logT_logn_diffgratings}
\end{figure*}

The fully Bayesian posterior density distributions of the plasma parameters from all the different datasets are shown in Figure~\ref{fig:logT_logn_diffgratings}.  Several characteristics of the results are apparent:
\hfil\break (1) The main contributor to the differences seen in the locations of the modes is calibration errors between the different instruments.  It is instructive to compare the magnitudes of these differences to the improvement achieved by accounting for atomic data errors, as illustrated in the widths of the posterior densities of the pragmatic and fully Bayesian methods shown in Figures~\ref{fig:fe17_ctnm_pragBfullB} and ~\ref{fig:ox_pragBfullB}.  For both \Fe\ and \OHe, $\Delta\logtemp\approx{0.2}$, and for \OHe, $\Delta\log\Ne\approx{0.4}$, comparable to the offsets between instruments, $\Delta\log\Te\approx{0.05-0.3}$ and $\Delta\log\Ne\approx{0.1-0.4}$.  We thus conclude that atomic data uncertainties contribute as much to the systematics of parameter estimates as instrument calibration uncertainties.
\hfil\break (2) In the \Fe\ results,
the posterior densities of $\logtemp$ are in distinct clusters, with {\tt am[p/n]} overlapping significantly with each other but well separated from {\tt al[p/n]}, which have a small but non-zero overlap.
These clusters straddle the nominal location of the spike in the Capella DEM at $\logtemp=6.8$, and are offset by $\approx-0.1,+0.2$~dex. Note that we expect systematic differences due to calibration uncertainties of as much as $\pm$10\% between the ACIS-S+HETG and ACIS-S+LETG configurations \citep{2021AJ....162..254M}, which is consistent with the estimated differences. Furthermore, this difference could also be due to attributed to intrinsic variability in Capella: the {\tt am[p/n]} data were obtained regularly over a variety of epochs (1999-aug, 2000-mar, 2001-feb, 2002-apr, 2003-sep, 2004-sep, 2005-mar, 2006-apr, 2008-apr, 2009-apr, 2009-nov, 2010-dec, 2011-dec, 2013-dec, 2014-dec, 2016-jul, 2018-dec) while the {\tt al[p/n]} data were obtained intermittently during only three epochs (1999-nov, 2007-apr, and the latter half of 2016) \citep[see][]{2022yCat..51620254M}; thus, some of the difference can be attributed to intrinsic differences in the high-energy emission in Capella \citep{1993ApJ...418L..41D}.
\hfil\break (3) In the \OHe\ $\logtemp$ estimates, the range of variation across datasets is smaller at $\approx{0.1}$~dex, and the modes of the posterior densities are lower than in the \Fe\ case.  The largest change is for {\tt al[p/n]} estimates, which go from $+0.2$~dex to $\approx-0.1$~dex relative to $\logtemp=6.8$. These differences from \Fe\ could be a consequence of our assumption of isothermality of the corona: the observed \OHe\ fluxes could be obtained from a lower-temperature DEM component than the \Fe\ fluxes.  While cognizant of the systematic shift this can produce between \Fe\ and \OHe\ estimates of $\logtemp$, the result from the latter is nevertheless robust, since in the two-stage analysis we allow the prior propagated from \Fe\ to have broader tails than the posterior density, and the \OHe\ estimates of $\logtemp$ are obtained independently.
\hfil\break (4) The estimated plasma densities are in the range ${\approx}9-25{\times}10^9$~cm$^{-3}$, with the individual posterior distributions offset by $\approx$0.2-0.3~dex from each other.  The range and offsets are again consistent with expected calibration uncertainties.
\hfil\break (5) The estimated densities suggest that the corresponding volume of emission is small, $\approx$10$^{31}$~cm$^3$. Most of the emission likely comes from the G8\,III component \citep[][]{2006ApJ...644L.117I,2023MNRAS.522L..66B} which has a radius of $\approx$12~R$_\odot$, and thus a total surface area of $\approx$10$^{25}$~cm$^2$. Thus, if the coronal emission is uniformly distributed over the surface, for an estimated volume $\volume\approx1.6{\times}10^{31}$~cm$^{3}$, its thickness is only $\approx$20~km, significantly smaller than the pressure scale height for 5~MK plasma, and is thus unrealistic. Therefore the emission must be patchy and concentrated in a few areas, and considering the lack of observed rotational modulation in Capella, must be located in the polar region.

\section{Summary}\label{sec:summary}

In this work, we have developed a statistical method to incorporate atomic data uncertainties while estimating the temperature and density of coronal plasma in Capella, using high-resolution and long-duration observations made using the \chandra\ gratings.

This represents the culmination of the second phase of our program to handle atomic uncertainties, following on from \citet{yu2018incorporating}, who developed a method to incorporate atomic emissivity uncertainties in estimating solar coronal plasma densities.  Here, we directly model the observed counts in selected lines from \Fe\ (Table~\ref{tab:fe17_adjusted_lrf}) and \OHe\ (Table~\ref{tab:ox_adjusted_lrf}) in Capella, obtained in various grating order and detector combinations (Section~\ref{sec:chandra_data}), using an isothermal approximation to the DEM.  Variants of atomic line emissivities of the selected lines are constructed as Monte Carlo samples: for \Fe, this is obtained by varying the collision strengths and decay rates randomly within a realistic uncertainty range, obtained by comparing different compilations (Section~\ref{sec:atomic_fe}); and for \OHe\ we first carry out ICFT {\sl R}-matrix calculations with different, randomly scaled parameters to alter the energy separations between levels and obtain different realizations of the rates and use them to construct the Monte Carlo emissivity samples (Section~\ref{sec:atomic_oxygen}).  These sample sets of randomly generated line emissivities are then decomposed using a principal components analysis, with up to 7 (for \Fe) and up to 6 (for \OHe) components kept which explain $\gtrsim$50\% of the variance (Section~\ref{sec:atomic_pca}).  This not only allows a dramatic reduction in the dimensionality of the atomic emissivity tables, but also allows a more fine-grained exploration of the space of atomic emissivity variations by effectively interpolating within the sample sets.  We carry out the analysis in two stages, first estimating the plasma temperature alone using the density-insensitive \Fe\ lines (Sections~\ref{sec:fe_model},\ref{sec:fe17stage}), and using the resulting posterior density to define an informative prior density for the second stage that uses the \OHe\ line fluxes to derive estimates of both plasma temperature and plasma density (Sections~\ref{sec:ox_model},\ref{sec:oxstage}).

We carry out analyses that accept the randomly generated atomic emissivities as unalterable as well as allow the data to select the better performing deviations among the emissivities (the {\sl pragmatic} and {\sl fully Bayesian} techniques respectively, see Section~\ref{sec:stats}).  We apply the method to the combined datasets of Capella obtained with \chandra\ between 1999 and 2020, obtained with both the $+$ve and $-$ve orders of ACIS-S/MEG ({\tt amp}, {\tt amn}) and ACIS-S/LEG ({\tt alp}, {\tt aln}).  Note that we do not use the HEG due to its lack of sufficient effective area over the \OHe\ range.  We find that the \Fe\ analysis estimates a coronal plasma temperature of $\approx$5~MK, with significant deviations required in the atomic emissivities based on the differences seen in the fully Bayesian posterior densities compared to the pragmatic Bayesian posterior densities (Figure~\ref{fig:fe17_ctnm_pragBfullB}).  The \OHe\ analysis reinforces the plasma temperature estimate of $\approx$3-5~MK, and further estimates a plasma density of $\approx{10}^{10}$~cm$^{-3}$, without requiring significant changes to the underlying atomic emissivities (Figure~\ref{fig:ox_pragBfullB}).  A plasma emission volume of ${\approx}1.5{\times}10^{31}$~cm$^{3}$ is required in both cases, suggesting that the coronal plasma is either spread thinly (the combined surface area of the binary is ${\approx}1.5{\times}10^{25}$~cm$^{2}$) or is localized.
A comparison of the results across the different spectral datasets shows that the results are consistent within expected calibration uncertainties (Figure~\ref{fig:logT_logn_diffgratings}).

The next phase in our program is to model functional forms of the DEM, with more than one parameter required to determine its shape as a function of temperature while incorporating atomic data uncertainties.  This will require the use of more species of lines, since otherwise the number of PCA components used to describe the uncertainties in atomic emissivity must be decreased, as otherwise the problem becomes overdetermined. Thus, attendant uncertainties in ion balance and composition would also have to be included in such an analysis.

\facilities 
This paper uses a list of \chandra\ datasets, obtained by the \chandra\ X-ray Observatory, contained in~\dataset[\chandra\ Data Collection (CDC) 228]{https://doi.org/10.25574/cdc.228}

\software {\sl CIAO} \citep[\url{https://cxc.harvard.edu/ciao/};][]{2006SPIE.6270E..1VF};
{\sl PINTofALE} \citep{2000BASI...28..475K}; {\sl CHIANTI} \citep{chianti_v10}; {\sl R} \citep[\url{https://cran.r-project.org};][]{R-citation}; {\sl Stan} \citep[\url{https://mc-stan.org};][]{Stan-citation}.

\section*{Acknowledgements}

%
%
%
%
%

This work was conducted under the auspices of the CHASC International Astrostatistics Center. CHASC is supported by NSF grants DMS-18-11308, DMS-18-11083, DMS-18-11661, DMS-21-13615, DMS-21-13397, and DMS-21-13605; by the UK Engineering and Physical Sciences Research Council [EP/ W015080/1]. We thank Mark Weber and Nathan Stein for many helpful discussions. V.L.K. further acknowledges support from NASA contract to the Chandra X-ray Center NAS8-03060. D.v.D. was also supported in part by Marie- Skłodowska-Curie RISE (H2020-MSCA-RISE-2015-691164, H2020-MSCA-RISE-2019-873089) Grants provided by the European Commission. G.D.Z. acknowledges support from Science and Technology Facilities Council UK (STFC) via the consolidated grants to the atomic astrophysics group at DAMTP, University of Cambridge (ST/P000665/1 and ST/ T000481/1). D.C.S. is supported by a Discovery Grant (RGPIN-2021-03985) provided by the National Sciences and Engineering Research Council of Canada. Finally, the authors acknowledge the generous support from the International Space Science Institute for hosting discussions among the ``Improving the Analysis of Solar and Stellar Observations'' international team. The full team was composed of Harry Warren (NRL; PI), Inigo Arregui (Inst. Astr. de Canarias), Frederic Auchere (Inst. Astr. Spatiale), Connor Ballance (Queen’s Univ.), David Stenning (IAP/Simon Fraser), Jessi Cisewski Kehe (Yale/ Wisconsin), Giulio Del Zanna (Cambridge), Veronique Delouille (Royal Obs. Belgium), Adam Foster (SAO), Chloe Guennou (Columbia), Vinay Kashyap (SAO; co-I), Fabio Reale (OAPA Palermo), Randall Smith (SAO), Nathan Stein (UPenn/Spotify), David A. van Dyk (Imperial), and Mark Weber (SAO; co-PI). 

\clearpage
\appendix 

\section{A Four-Step Gibbs Sampler}
\label{sec:app-algs}

In this section we describe the Gibbs sampler used to obtain Monte Carlo samples from the posterior distributions given in Section~\ref{sec:emiss-uncerntainty}. The Gibbs sampler is used to cross check and verify numerical results obtained with Stan.

To derive this Gibbs sampler we write the posterior distribution in terms of the (unobserved) background counts, $\dataso_{\rm B}(\wvl)$, that contaminate the (observed) source counts, $\dataso(\wvl)$, where as in Section~\ref{sec:notation}, $\wvl$ indexes the recorded photon wavelength.  We denote the collection of contaminating background counts by $\dataso_{\rm B} =\{\dataso_{\rm B}(\wvl), \wvl\in\mathcal{W}\}$. \cite{yu2020multistage} shows that we can obtain a Monte Carlo sample from the target posterior,
$
    p(\dataso_{\rm B}, \theta_{\rm S},\theta_{\rm B},\rv \mid \dataso, \databg),
$
via the Four-step Gibbs Sampler the implements the following steps at iteration $(i)$

\paragraph{Step 1} For each $\wvl$ in $\cal W$, sample 
$$
Y_B(\wvl)^{(i)} \mid Y(\wvl), \theta_{\rm S}^{(i-1)}, \theta_{\rm B}^{(i-1)}, \rv^{(i-1)} \sim \operatorname{Binomial} \left( Y(\wvl), \frac{\theta_{\rm B}}{\theta_{\rm B}+s(\wvl, \theta_{\rm S}, \rv) + \kappa(\wvl, \theta_{\rm S})} \right)
$$
where $\theta_{\rm S}^{(i-1)} = (\log n^{(i-1)}, \log \volume^{(i-1)}, \logtemp^{(i-1)})$

\paragraph{Step 2} Sample 
$$
\theta_{\rm B}^{(i)} \mid Z, Y_{\rm B}^{(i)} \sim \operatorname{Gamma} \left( 
\sum_{j=1}^H Z(\wvl_j) + \sum_{j=1}^H Y_B(\wvl_j) + \alpha_B,
H(\ratio+1) + \beta_B \right),
$$ 
where we specify the prior distribution 
$\theta_{\rm B} \sim \operatorname{Gamma} \big(\alpha_B = \hbox{shape} = 0.5, \beta_B= \hbox{rate} = 2 \big)$ in Section~\ref{sec:priors}. 

\paragraph{Step 3} Update $\rv^{(i)}$. For the pragmatic-Bayesian methods this simply involves sampling from the prior distributions given in Equations~\ref{eq:prior_rvfe} and \ref{eq:prior_rvox} for the \Fe\ and \OHe\ analyses, respectively. For the fully-Bayesian methods, we update $\rv^{(i)}$ via a Metropolis step 
\citep{metropolis1953equation,hastings1970monte} with an adaptive jumping rule
\citep{rosenthal2011optimal}, as described in Equations~(5.36)-(5.39) of \cite{yu2020multistage} for the \Fe\ analysis, and analogously for the  \OHe\ analysis. 

\paragraph{Step 4} Update $\theta_S^{(i)}$ via a Metropolis step with an adaptive jumping rule, as described in Equations~(5.32)-(5.35) of \cite{yu2020multistage} for the \Fe\ analyses and Equations~(6.31)-(6.33) of \cite{yu2020multistage} for the \OHe\ analyses. 

For the first 1000 iterations, the adaptive jumping rules in Steps~3 and 4 are $d$-dimensional multivariate normal distributions with means set equal to the current value of the corresponding parameter and with variance-covariance matrix set equal to $0.1^2 \Sigma_0/d$, where $\Sigma_0$ is a diagonal matrix with diagonal entries equal to the prior variances of the corresponding parameters.
The initial 1000 draws are used to tune  the adaptive Metropolis. Specifically, starting with iteration 1001, we follow \citet{roberts2009examples} and use a mixture of two $d$-dimensional multivariate Gaussian distributions for the jumping rules. Both multivariate Gaussian distributions are centered at the current value of the parameter being updated, one with variance-covariance matrix equal to that of the diagonal scaled prior used in the initial draws 
and the other matrix set equal to 
$2.38^2\Sigma/d$, where $\Sigma$ is the empirical variance-covariance matrix of the full history of the chain.
The mixture component corresponding to the prior distribution is given weight 0.05 and the empirical variance-covariance matrix of the second component is updated every fiftieth iteration to reduce computational costs.  

\section{Comparison of algorithms and output data analysis}
\label{app:compare_algs}

\begin{figure}[!htbp]
\centering
\includegraphics[width=0.53\linewidth]{ 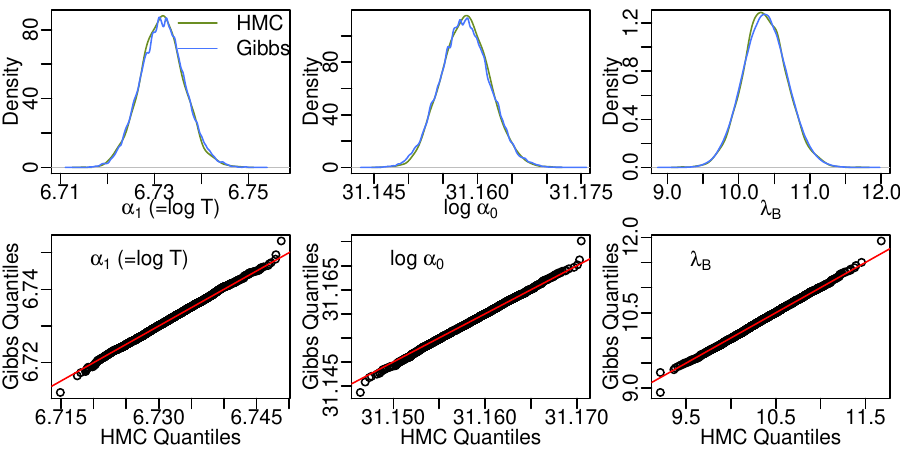}
\includegraphics[width=0.56\linewidth]{ 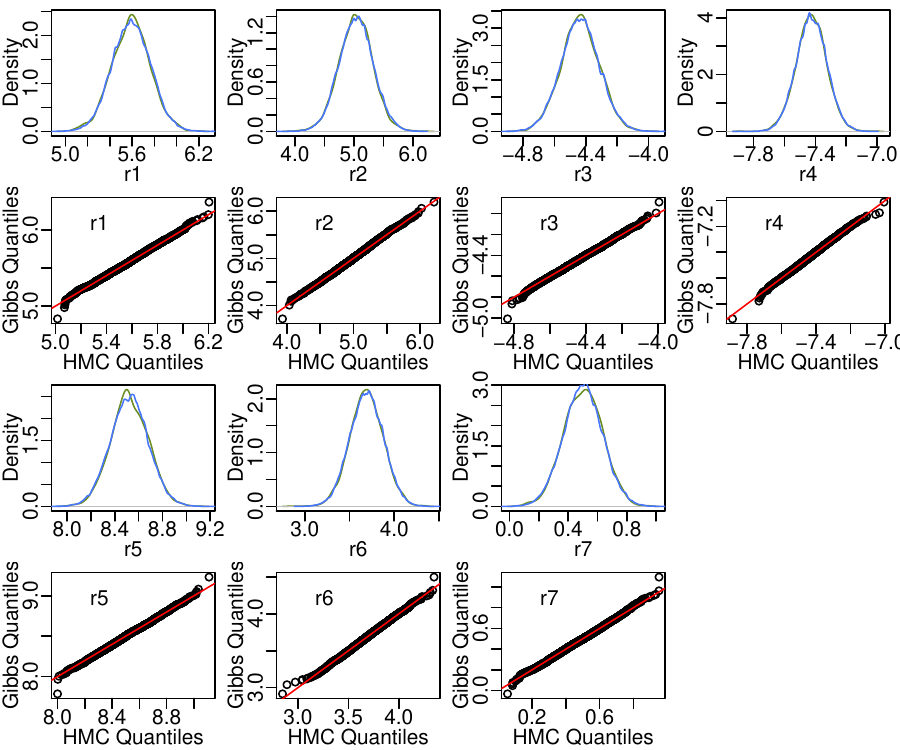}
\caption{Comparison of the MC samples obtained with the four-step Gibbs Sampler (blue) and the HMC (green) deployed for a fully-Bayesian analysis of the \Fe\ Capella line counts. The panels compare the  marginal histograms and provide quantile-quantile plots of the samples. The 45-degree lines (equal quantiles) are plotted in red. The two samplers give consistent results.}
\label{fig:fe17_ctnm_4StepGibbsHMC}
\end{figure}

We consider two algorithms that can be used to obtain a MC sample from the fully-Bayesian posterior distribution, the four-step Gibbs Sampler and HMC via Stan. As a diagnostic, we compared the results of both and found good alignment of the marginal distributions of each of the model parameters in the \Fe\ analysis described in Section~\ref{sec:fe_model}, see Figure~\ref{fig:fe17_ctnm_4StepGibbsHMC}.
In our numerical results, HMC exhibits much lower autocorrelations (along the Markov chain) and thus required a much smaller MC sample which can be obtained much more quickly. Thus HMC approach is preferred. 

\section{LRF parameter estimates of other line-grating combinations}
\label{app:LRF_other_data_sets}

The tables~\ref{tab:fe17_adjusted_lrf_amn}-\ref{tab:ox_adjusted_lrf_aln} present the minimum $\chi^2$ estimates of the location and scale parameters, $\omega$ and $\sigma$, of the LRF given in Equation~\ref{eq:lrf} for \Fe\ and \OHe\ spectral lines of other line-grating combinations, {\tt amn}, {\tt alp}, and {\tt aln}. The corresponding effective areas are also provided.

\begin{table}
\begin{center}
\caption{A summary of the best-fit scale, $\hat{\sigma}$, best-fit location, $\hat{\ww}$, and nominal location, $\ww$, of LRF for each \Fe \ {\tt amn} spectral line.}
\begin{tabular}{c c c c } 
 \hline
 \hline
$\ww \ (\mathring{\text{A}})$ & $\hat{\ww} \ (\mathring{\text{A}})$ &  $\hat{\sigma}$  & effective area $(\text{cm}^2 k\text{sec})$  \\
 \hline
$11.129$ & $11.131$ & $0.0080$ & $15340$  \\
$11.250$ & $11.252$ & $0.0080$ & $14750$  \\
$12.124$ & $12.132$ & $0.0090$ & $11170$  \\
$12.264$ & $12.266$ & $0.0090$ & $10730$  \\
$13.825$ & $13.826$ & $0.0105$ & $5302$  \\
$13.890$ & $13.892$ & $0.0095$ & $5399$ \\
$15.013$ & $15.014$ & $0.0090$ & $7737$  \\
$15.262$ & $15.262$ & $0.0085$ & $7274$  \\
$15.453$ & $15.453$ & $0.0125$ & $6920$  \\
$16.336$ & $16.334$ & $0.0085$ & $5557$  \\
$16.776$ & $16.776$ & $0.0080$ & $4895$  \\
$17.051$ & $17.051$ & $0.0080$ & $4537$  \\
$17.096$ & $17.096$ & $0.0075$ & $4508$  \\
\hline\hline
\end{tabular}
\label{tab:fe17_adjusted_lrf_amn}
\end{center}
\end{table}

\begin{table}
\begin{center}
\caption{A summary of the best-fit scale, $\hat{\sigma}$, best-fit location, $\hat{\ww}$, and nominal location, $\ww$, of LRF for each \Fe \ {\tt alp} spectral line.}
\begin{tabular}{c c c c } 
 \hline
 \hline
$\ww \ (\mathring{\text{A}})$ & $\hat{\ww} \ (\mathring{\text{A}})$ &  $\hat{\sigma}$  & effective area $(\text{cm}^2 k\text{sec})$  \\
 \hline
$11.129$ & $11.132$ & $0.0130$ & $4428$  \\
$11.250$ & $11.250$ & $0.0500$ & $4371$  \\
$12.124$ & $12.128$ & $0.0160$ & $4064$  \\
$12.264$ & $12.267$ & $0.0200$ & $4026$  \\
$13.825$ & $13.827$ & $0.0170$ & $3671$  \\
$13.890$ & $13.896$ & $0.0210$ & $3654$ \\
$15.013$ & $15.010$ & $0.0180$ & $3365$  \\
$15.262$ & $15.258$ & $0.0165$ & $3300$  \\
$15.453$ & $15.449$ & $0.0210$ & $3245$  \\
$16.336$ & $16.326$ & $0.0310$ & $2966$  \\
$16.776$ & $16.774$ & $0.0155$ & $2824$  \\
$17.051$ & $17.049$ & $0.0170$ & $2740$  \\
$17.096$ & $17.093$ & $0.0140$ & $2733$  \\
\hline\hline
\end{tabular}
\label{tab:fe17_adjusted_lrf_alp}
\end{center}
\end{table}

\begin{table}
\begin{center}
\caption{A summary of the best-fit scale, $\hat{\sigma}$, best-fit location, $\hat{\ww}$, and nominal location, $\ww$, of LRF for each \Fe \ {\tt aln} spectral line.}
\begin{tabular}{c c c c } 
 \hline
 \hline
$\ww \ (\mathring{\text{A}})$ & $\hat{\ww} \ (\mathring{\text{A}})$ &  $\hat{\sigma}$  & effective area $(\text{cm}^2 k\text{sec})$  \\
 \hline
$11.129$ & $11.133$ & $0.0155$ & $3236$  \\
$11.250$ & $11.250$ & $0.0500$ & $3160$  \\
$12.124$ & $12.132$ & $0.0145$ & $2849$  \\
$12.264$ & $12.271$ & $0.0165$ & $2774$  \\
$13.825$ & $13.828$ & $0.0175$ & $1987$  \\
$13.890$ & $13.900$ & $0.0305$ & $2035$ \\
$15.013$ & $15.017$ & $0.0170$ & $1886$  \\
$15.262$ & $15.265$ & $0.0170$ & $1723$  \\
$15.453$ & $15.449$ & $0.0175$ & $1714$  \\
$16.336$ & $16.333$ & $0.0500$ & $1474$  \\
$16.776$ & $16.778$ & $0.0150$ & $1338$  \\
$17.051$ & $17.054$ & $0.0160$ & $1266$  \\
$17.096$ & $17.098$ & $0.0145$ & $1258$  \\
\hline\hline
\end{tabular}
\label{tab:fe17_adjusted_lrf_aln}
\end{center}
\end{table}

\begin{table}
\begin{center}
\caption{A summary of the best-fit scale $\hat{\sigma}$, best-fit location, $\hat{\ww}$, and nominal location, $\ww$, of LRF for each \OHe \ {\tt amn} spectral line.}
\begin{tabular}{c c c c } 
 \hline
 \hline
$\ww \ (\mathring{\text{A}})$ & $\hat{\ww} \ (\mathring{\text{A}})$ &  $\hat{\sigma}$  & effective area $(\text{cm}^2 k\text{sec})$  \\
 \hline
$17.396$ & $17.3990$ & $0.0127$ & $4035$  \\
$17.768$ & $17.7804$ & $0.0223$ & $3618$  \\
$18.627$ & $18.6299$ & $0.0108$ & $3257$  \\
$21.602$ & $21.6029$ & $0.0103$ & $1565$  \\
$21.805$ & $21.8055$ & $0.0084$ & $1446$  \\
$22.101$ & $22.0995$ & $0.0085$ & $1294$  \\
 \hline
 \hline
\end{tabular}
\label{tab:ox_adjusted_lrf_amn}
\end{center}
\end{table}

\begin{table}
\begin{center}
\caption{A summary of the best-fit scale $\hat{\sigma}$, best-fit location, $\hat{\ww}$, and nominal location, $\ww$, of LRF for each \OHe \ {\tt alp} spectral line.}
\begin{tabular}{c c c c } 
 \hline
 \hline
$\ww \ (\mathring{\text{A}})$ & $\hat{\ww} \ (\mathring{\text{A}})$ &  $\hat{\sigma}$  & effective area $(\text{cm}^2 k\text{sec})$  \\
 \hline
$17.396$ & $17.3998$ & $0.0146$ & $2659$  \\
$17.768$ & $17.7727$ & $0.0161$ & $2499$  \\
$18.627$ & $18.6313$ & $0.0177$ & $2470$  \\
$21.602$ & $21.5963$ & $0.0210$ & $1037$  \\
$21.805$ & $21.7986$ & $0.0130$ & $878.3$  \\
$22.101$ & $22.0945$ & $0.0162$ & $538.1$  \\
 \hline
 \hline
\end{tabular}
\label{tab:ox_adjusted_lrf_alp}
\end{center}
\end{table}

\begin{table}
\begin{center}
\caption{A summary of the best-fit scale $\hat{\sigma}$, best-fit location, $\hat{\ww}$, and nominal location, $\ww$, of LRF for each \OHe \ {\tt aln} spectral line.}
\begin{tabular}{c c c c } 
 \hline
 \hline
$\ww \ (\mathring{\text{A}})$ & $\hat{\ww} \ (\mathring{\text{A}})$ &  $\hat{\sigma}$  & effective area $(\text{cm}^2 k\text{sec})$  \\
 \hline
$17.396$ & $17.3957$ & $0.0131$ & $1115$  \\
$17.768$ & $17.7743$ & $0.0184$ & $1053$  \\
$18.627$ & $18.6348$ & $0.0189$ & $969.9$  \\
$21.602$ & $21.6009$ & $0.0141$ & $461.9$  \\
$21.805$ & $21.8049$ & $0.0136$ & $415.6$  \\
$22.101$ & $22.0981$ & $0.0092$ & $339.7$  \\
 \hline
 \hline
\end{tabular}
\label{tab:ox_adjusted_lrf_aln}
\end{center}
\end{table}

\newpage

\bibliography{mybibl}
\bibliographystyle{aasjournal}

\end{document}